\documentclass[longauth]{aa}

\pdfoutput=1

\makeatletter
\newcommand{\lam}{$\lambda$ }
\newcommand{\lamb}{$\lambda$}
\newcommand{\qp}{$Q_\phi$ }
\newcommand{\up}{$U_\phi$ }
\newcommand{\rxj}{RX\,J1615 }
\newcommand{\rxjb}{RX\,J1615}
\newcommand{\thwp}{$\theta_\mathrm{hwp}$ }
\newcommand{\ff}{f\mbox{}f}
\newcommand{\ffs}{f\mbox{}f }
\newcommand{\ffi}{f\mbox{}f\mbox{i}}
\newcommand{\fix}{f\mbox{}i}
\newcommand{\fl}{f\mbox{}l}

\newcommand{\aib}{{$A1$}}
\newcommand{\aii}{{$A2$} }

\newcommand{\ri}{{$R1$} }
\newcommand{\rib}{{$R1$}}
\newcommand{\rii}{{$R2$} }
\newcommand{\riib}{{$R2$}}
\newcommand{\ii}{{$I1$} }
\newcommand{\iib}{{$I1$}}

\newcommand{\snew}{\rm}
\newcommand{\enew}{\rm}

\usepackage[english]{babel}
\usepackage{graphicx}
\usepackage[breaklinks=true]{hyperref} 
\usepackage{multicol}
\usepackage{natbib,twoopt}
\usepackage[varg]{txfonts}
\bibpunct{(}{)}{;}{a}{}{,} 	

\hypersetup{colorlinks=true,citecolor=blue}

\begin{document} 
\title{Multiple rings in the transition disk and companion candidates around RX\,J1615.3-3255.} 
\subtitle{High contrast imaging with VLT/SPHERE.\thanks{
Based on observations made with ESO Telescopes at the La Silla Paranal Observatory under programme IDs
095.C-0298(A), 095.C-0298(B), and 095.C-0693(A) during guaranteed and open time observations of the SPHERE consortium, 
and on NACO observations: program IDs: 085.C-0012(A), 087.C-0111(A), and 089.C-0133(A).
}
}
\titlerunning{SPHERE imaging of the disk and companion candidates around \rxjb.}

\author{
J.~de Boer\inst{1} 
\and G.~Salter \inst{2} 
\and M.~Benisty \inst{3} 
\and A.~Vigan \inst{2} 
\and A.~Boccaletti \inst{4} 
\and P.~Pinilla \inst{1} 
\and C.~Ginski \inst{1} 
\and A.~Juhasz \inst{5} 
\and A.-L.~Maire \inst{6} 
\and S.~Messina \inst{7} 
\and S.~Desidera \inst{8} 
\and A.~Cheetham \inst{9} 
\and J.\,H.~Girard \inst{10} 
\and Z.~Wahhaj \inst{10} 
\and M.~Langlois \inst{11}\inst{, 2}
\and M.~Bonnefoy \inst{3}
\and J.-L.~Beuzit \inst{3}
\and E.~Buenzli \inst{6}
\and G.~Chauvin \inst{3}
\and C.~Dominik \inst{12}
\and M.~Feldt \inst{6}
\and R.~Gratton \inst{7}
\and J.~Hagelberg \inst{9}
\and A.~Isella \inst{13}
\and M.~Janson \inst{14}
\and C.\,U.~Keller \inst{1}
\and A.-M.~Lagrange \inst{3}
\and J.~Lannier \inst{3}
\and F.~Menard \inst{3}
\and D.~Mesa \inst{7}
\and D.~Mouillet \inst{3}
\and M.~Mugrauer\inst{15}
\and S.~Peretti \inst{9}
\and C.~Perrot \inst{4}
\and E.~Sissa \inst{7}
\and F.~Snik \inst{1}
\and N.~Vogt\inst{16}
\and A.~Zurlo \inst{17} \inst{, 2}
\and SPHERE Consortium
} 
\institute{
Leiden Observatory, Leiden University, P.O. Box 9513, 2300RA Leiden, The Netherlands. \\
\email{deboer@strw.leidenuniv.nl} 
\and Aix Marseille Universit\'e, CNRS, LAM (Laboratoire d'Astrophysique de Marseille) UMR 7326, 13388, Marseille, France
\and Universit\'e Grenoble Alpes, IPAG, F-38000 Grenoble, France
\and LESIA, CNRS, Observatoire de Paris, Universit\'e  Paris Diderot, UPMC, 5 place J. Janssen, 92190 Meudon, France.
\and Institute of Astronomy, Madingley Road, Cambridge CB3 OHA, UK.
\and Max-Planck-Institut fuer Astronomie, Koenigstuhl 17, 69117 Heidelberg, Germany
\and INAF – Osservatorio Astronomico di Padova, Vicolo dell’Osservatorio 5, 35122 Padova, Italy.
\and INAF Catania Astrophysical Observatory, via S. So\fix a 78, 95123 Catania, Italy.
\and Observatoire de Gen\`eve, Universit\'e de Gen\`eve, 51 Chemin des Maillettes, 1290, Versoix, Switzerland. 
\and European Southern Observatory, Alonso de Cordova 3107, Casilla 19001 Vitacura, Santiago 19, Chili.
\and Observatoire de Lyon, Centre de Recherche Astrophysique de Lyon, Ecole Normale Sup\'erieure de Lyon, 
CNRS, Universit\'e  Lyon 1, UMR 5574, 9 avenue Charles Andr\'e, Saint-Genis Laval, 69230, France
\and Sterrenkundig Instituut Anton Pannekoek, Science Park 904, 1098 XH Amsterdam, The Netherlands
\and Department of Physics \& Astronomy, Rice University, 6100 Main Street, Houston, TX 77005, USA
\and Department of Astronomy, Stockholm University, AlbaNova University Center, 106 91 Stockholm, Sweden
\and Astrophysical Institute and University Observatory Jena, Schillerg\"a\ss chen 2, 07745 Jena, Germany
\and Instituto de F\'isica y Astronom\'ia, Facultad de Ciencias, Universidad de Valpara\'iso, Av. Gran Breta\~na 1111, Playa Ancha, Valparaíso, Chile.
\and N\'ucleo de Astronom\'ia, Facultad de Ingenier\'ia, Universidad Diego Portales, Av. Ejercito 441, Santiago, Chile
}

\date{Received July 8, 2016; accepted September 22, 2016}

\abstract
{The e\ff ects of a planet sculpting the disk from which it formed are most likely to be found in
disks in transition between classical protoplanetary and debris disk.
Recent direct imaging of transition disks has revealed structures such as dust rings, gaps, and spiral arms, but an unambiguous link between these structures and sculpting planets is yet to be found. 
}
{
We search for signs of ongoing planet-disk interaction 
and study the distribution of small grains at the surface
of the transition disk around RX\,J1615.3-3255 (\rxjb).
}
{
We observed \rxj with VLT/SPHERE:
we obtained polarimetric imaging 
with ZIMPOL ($R'$-band) and IRDIS ($J$);
and IRDIS ($H2H3$) dual-band imaging 
with simultaneous spatially resolved spectra with the IFS ($YJ$).
}
{
We image the disk for the \fix rst time in scattered light and detect two arcs, two rings, a gap and an inner disk with marginal evidence for an inner cavity.
The shapes of the arcs suggest that they probably are segments of full rings. 
Ellipse \fix tting for the two rings and inner disk yield a disk inclination $i = 47 \pm 2^\circ$ and \fix nd semi-major axes of $1.50 \pm 0.01''$ (278\,au), $1.06 \pm 0.01''$ (196\,au) and $0.30 \pm 0.01''$ (56\,au), respectively.
We determine the scattering surface height above the midplane, 
based on the projected ring center o\ff sets.
Nine point sources are detected between 2.1$''$ and  8.0$''$ separation and considered as companion candidates. 
With NACO data we recover four of the nine point sources, which we determine not to be co-moving, and therefore unbound to the system.
}
{
We present the \fix rst detection of the transition disk of \rxj in scattered light.
The height of the rings indicate limited \fl aring of the disk surface, 
which enables partial self-shadowing in the disk.
The outermost arc either traces the bottom of the disk 
or it is another ring with semi-major axis $\gtrsim 2.35''$ (435\,au). 
We explore both scenarios, extrapolating the complete shape of the feature,
which will allow to distinguish between the two in future observations.
The most interesting scenario, where the arc traces the bottom of the outer ring, requires the disk truncated at $r \approx 360$\,au. 
The closest companion candidate, if indeed orbiting the disk at $540$\,au, would then be the most likely cause for such truncation. 
This companion candidate, as well as the remaining four,
require follow up observations to determine if they are bound to the system.
}

      \maketitle
      
%

 \begin{table*}
	\resizebox{0.785\textwidth}{!}
	{\begin{minipage}{\textwidth}   	
	\begin{tabular}{  c | c | c| c | c | c|c|r |r |c|c |c|}
	 	Date             & Instrument & Mode  & Coronagraph  & \fix lter  & $\lambda_0(\mu$m) & FWHM$^{1}$ (nm) & DIT (s) &$t_\mathrm{tot}$ (min) 
	 	& Seeing ($''$)& SR$^{2}$(\%) & FWHM$^{3}$ (mas)
	 	\\ \hline \hline
	       12-05-2015 & IRDIS          & IRDIFS & ALC\_YJH\_S  & $H2H3$  & 1.59 \& 1.67  &53 \& 55  & 64 & 136.5 & 0.65 & 65-70 & 44.5 \& 46.6\\ 
	       12-05-2015 & IFS              & IRDIFS & ALC\_YJH\_S  & $YJ$       &0.96 - 1.34    & 55.1 & 64 &136.5  & 0.65 & 45-65 & 30.0 - 36.1\\ 
	       15-05-2015 & IRDIS          & IRDIFS & ALC\_YJH\_S  & $H2H3$  & 1.59 \& 1.67  &53 \& 55 &  64 &68.3  & 0.65 & 65-70 & 44.5 \& 46.6\\ 
	       15-05-2015 & IFS              & IRDIFS & ALC\_YJH\_S  & $YJ$       &0.96 - 1.34    & 55.1    &  64 &68.3 & 0.65 & 45-65 & 30.0 - 36.1 \\ 
	       06-06-2015 & IRDIS          &  DPI & ALC\_YJH\_S  & $J$    & 1.26              & 197 &  64 & 76.8 & 1.3 &30-35 & 42.5 \\ 
	       09-06-2015 & ZIMPOL      & P2 & -                             & $R'$    & 0.626            & 149 & 120 &96  & 1.0 & $< 3$ & 95.5\\ 
 \hline
	\end{tabular}	 
	\vspace{1mm}
	\end{minipage}}
	 \caption{SPHERE observations of \rxjb.
	 $^{1}$ Full Width at Half Maximum of the \fix lter transmission. For the IFS, the FWHM is given per spectral channel.
	$^{2}$ Strehl Ratios are measured at the observing wavelength using the unsaturated PSFs (Girard et al. in prep.). 
	For the low level of AO correction obtained for ZIMPOL, the FWHM is the appropriate image quality metric.	 
	 $^{3}$ Full Width at Half Maximum of the PSF. }

		\label{tab:obs}
\end{table*}

\section{Introduction} \label{sec:intro}

  	The evolution of circumstellar disks around pre-main sequence stars, 
  	and planet-disk interactions
  	constitute two of the major components in our understanding of planet formation.
  	Determining the disk structure (such as its geometry, gas and dust distribution) is essential to our
  	understanding of both disk evolution and planet-disk interactions.
	Direct imaging of protoplanetary disks shows a variety of features, such as inner cavities, gaps, 
	rings and spirals that can be observed both in early \citep[HL Tau; $\le 1$\,Myr,][]{ALMA:2015} 
	and late evolutionary phases \citep[e.g HD100546, $\sim10$\,Myr,][]{Grady:2001}.
  	This poses the questions on whether the shapes and features of protoplanetary disks are caused by the presence of planets, 
 	or if the disk structures are regulating the formation of planets, or both.
	If the disk structures are not created by nearby planets, we still need to determine what did create these structures.

\citet{Strom:1989AJ} classi\fix ed a subset of protoplanetary disks by a dip	
	in the InfraRed (IR) range of their Spectral Energy Distribution (SED),
	caused by an inner dust depleted region (cavity).
	The authors suggested these disks represent an evolutionary stage `in transition' between
	classical (younger) protoplanetary disks and more evolved debris disks.
	Multiple explanations have been suggested for the origin of cavities within transition disks, 
	such as photo-evaporation \citep[e.g.][]{Hollenbach:2005prpl}, 
	and clearing by massive planets \citep[e.g.][]{Strom:1989AJ, Alexander:2009ApJ, Pinilla:2015a}.
	Transition disks also display other potential tracers of planet-disk interaction: the large
	scale structures 
	in disks such as spiral arms in e.g. MWC 758 \citep{Grady:2013ApJ,Benisty:2015A&A}; 
	ring structures, as seen in e.g. RX J1604-2130 \citep{Mayama:2012ApJ,Pinilla:2015a} 
	and TW\,Hydrae \citep[van Boekel et al.~submitted;]{Andrews:2016}; 
	and dust trapping vortices in e.g. IRS 48 \citep{Marel:2013Sci}.

	RX\,J1615.3-3255, or 2MASS\,J16152023-3255051 (hereafter \rxjb) has previously been determined to be a 1.1\,M$_\odot$ 
	pre-main sequence star with an age of 1.4~Myr 
	\citep{Wahhaj:2010ApJ},
	and a member of the Lupus star forming region at a distance d\,$= 185$\,pc \citep{Krautter:1997A&AS}. 
	From its very strong H$_{\alpha}$ emission  
	\citet{Wahhaj:2010ApJ} determine that \rxj is a Classical T-Tauri Star (CTTS), and
	classify this target as having an IR-excess with a 
	`Turn-on wavelength (\lamb) in the range (3.6 - 8\,$\mu$m) of the IR Array Camera' (TIRAC, 
	where turn-on \lam is the smallest \lam with disk emission distinguishable from stellar emission).
	The authors describe TIRAC targets as young ($1.3~\pm~0.3$\,Myr) objects with a small inner clearing in the disks, 
	i.e. transition disks.
	\citet{Andrews:2011ApJ} have resolved the disk with the SMA interferometer at 880$\,\mu$m, 
	determined a disk Position Angle \textit{PA} $ = 143^\circ$, inclination $i = 41^\circ$ 
	and found in their best-fit model a low-density cavity in the dust disk within $r_\mathrm{cav} = 30$\,au.
	The disk is resolved with ALMA (cycle 0) in
	$^{12}$CO (6-5) and 690 GHz continuum data by \citet{Marel:2015A&A}.
	The authors \fix nd \textit{PA} $ = 153^\circ, i = 45 \pm 5^\circ$ and from their model a dust cavity inside $r_\mathrm{cav} = 20$\,au.
	\snew
	Both \citet{Andrews:2011ApJ} and \citet{Marel:2015A&A} have used a gas-to-dust mass ratio of 100:1 in their models for the outer disk,
	and \fix nd a total disk mass of $0.128$\,M$_\odot$ and $0.474$\,M$_\odot$, respectively.
	\enew

	While sub-mm observations probe emission of the mm-sized grains in the deeper layers of the disk, 
	the surface of the micron sized dust disk can be detected by its scattering of Near-IR starlight. 
	The large contrast between starlight and disk surface brightness at NIR wavelengths requires us to remove the stellar speckle halo, 
	in order to detect the light scattered by the disk.

	Since 2014 
	the Spectro-Polarimetric High-contrast Exoplanet REsearch \citep[SPHERE,][]{Beuzit:2008SPIE} 
	instrument has been commissioned at the Very Large Telescope (VLT).
	SPHERE is an extreme adaptive optics (AO) assisted instrument designed for high-contrast imaging of 
	young giant exoplanets and circumstellar disks. 

	We use SPHERE to image the disk of \rxj in scattered light at multiple wavelengths with the aim of constraining the 3D disk geometry.
	In addition, we perform a search for possible companions that could be responsible for sculpting the disk.
	We report the new observations and the data reduction in Sections~\ref{sec:obsred} and \ref{sec:reduce}, respectively.
	Section~\ref{sec:results} describes the new detections of the disk structures and companion candidates.
	In Section~\ref{sec:discuss}, we give constraints on the disk geometry and discuss possible scenarios for the disk vertical structure. 
	We list our conclusions in Section~\ref{sec:conclude}.
	Based on archival data, we study the stellar properties of \rxj in Appendix~\ref{sec:stellar2}.

\section{Observations}
\label{sec:obsred}
\subsection{VLT/SPHERE IRDIS, IFS and ZIMPOL}

	We observed \rxj at multiple wavelengths using di\ff erent modes of SPHERE (listed in Table~\ref{tab:obs}).
	The extreme AO system, SAXO \citep[SPHERE AO for eXoplanet Observation,][]{Fusco:2014eo} includes a 41$\times$41-actuator deformable mirror, 
	pupil stabilization, di\ff erential tip tilt control and stress polished toric mirrors \citep{Hugot2012A&A} 
	for beam transportation to the coronagraphs 
	\citep{Boccaletti2008SPIE, Martinez:2009}
	and science instruments. 
	The latter comprise a near-InfraRed Dual-band Imager and Spectrograph \citep[IRDIS,][]{Dohlen:2008SPIE}, 
	a near-infrared Integral Field Spectrometer \citep[IFS,][]{Claudi:2008SPIE}, 
	and the Zurich IMaging POLarimeter \citep[ZIMPOL,][]{Thalmann:2008SPIE}.

	The SpHere INfrared survey for Exoplanets (SHINE), 
	executed during the Guaranteed Time Observations (GTO) of the SPHERE consortium, 
	includes observations of \rxj on May 12 and 15 of 2015.
	The observations were performed with the IRDIFS mode \citep{Zurlo:2014A&A} in pupil tracking, using 
	IRDIS in dual-band imaging mode \citep{Vigan:2010MNRAS} in the $H2H3$-bands 
	and IFS in $YJ$-band. 
	The observations are taken with the Apodized pupil Lyot Coronagraph ALC\_YJH\_S, 
	which has a diameter of 185\,mas.
	During the observations of May 12
	the \fix eld rotated by $113.2^\circ$. 
	On May 15, we observed the system with 
	a \fix eld rotation of $75.3^\circ$.

	On June 6, 2015, the SPHERE-Disk GTO program has observed \rxj
	with 	IRDIS in Dual-band Polarimetric Imaging mode \citep[DPI,][de Boer et al., in prep.]{Langlois:2014} 
	in $J$-band with the ALC\_YJH\_S coronagraph.
	For the six polarimetric cycles, we observed with the Half Wave Plate (HWP) at the angles 
	\thwp $= 0^\circ, 45^\circ, 22.5^\circ$, and $67.5^\circ$ to modulate the linear Stokes parameters ($Q$ and $U$). 
	Per \thwp we have taken three exposures.

	During open time observations on June 9, 2015 we have observed \rxj 
	with SPHERE's visible light polarimeter, ZIMPOL in \fix eld tracking (P2) non-coronagraphic mode in 
	$R'$-band with the dichroic beamsplitter.
	We have recorded six polarimetric (or HWP) cycles, with $\theta_\mathrm{hwp} = 0^\circ$,$ 45^\circ$,
	$22.5^\circ$, and $67.5^\circ$ to modulate $Q$ and $U$. 
	For each \thwp two frames were recorded for each of the two detectors of ZIMPOL. 
		
	The faint guide star \citep[$R = 11.21$\,mag,][]{Makarov:2007ApJ} poses a challenge for the AO system. 
	Especially in the visible light which is split between ZIMPOL and the SAXO wave front sensor, the moderate seeing ($\sim$ 1.0$''$) 
	yielded very poor Strehl ratio SR$_{R'} < 3\%$ and FWHM$_{R'}= 95.5 \pm 1.5$\,mas.

	\subsection{VLT/NACO and Keck/NIRC2 SAM}
									
	Additionally, \rxj was observed before with VLT/NACO 
	\citep{Lenzen:2003iu,Rousset:2003hh} 
	as part of a survey for substellar companions in the Lupus star forming region 
	during three epochs on May 7th 2010, May 8th 2011 and August 7th 2012 (Mugrauer et al. in prep.).
	The system was imaged in the $K_s$-band ($\lambda_0 = 2.18$\,$\mu$m, $\Delta \lambda = 350$\,nm) in 
	standard jitter imaging (\fix eld stabilized) mode.
	NACO's cube mode was used to save each individual frame \citep{Girard:2010}. 
	The Detector Integration Time (DIT) for a single exposure was 1\,s for the three epochs

	To investigate the possibility of stellar or high-mass substellar companions at close separations, 
	we include the results from archival Keck NIRC2 Sparse Aperture Masking (SAM) data. 
	\rxj was observed on
	April 4, 2012 with $\mathrm{DIT} = 5$\,s, 4 coadds per frame, 12 frames in total; 
	on July 8, 2012 with $\mathrm{DIT} = 5$\,s, 4 coadds per frame, 62 frames in total; 
	and on June 10, 2014 with $\mathrm{DIT} = 10$\,s, 2 coadds per frame, 40 frames in total. 
	During all three epochs, the $K'$-\fix lter and the nine hole mask have been used.

\section{Data Reduction}
\label{sec:reduce}
    
\subsection{ZIMPOL P2 observations in $R'$-band}
    \begin{figure*}
   \centering
   \includegraphics[width=1.1\textwidth, trim = 50 0 -50 10]{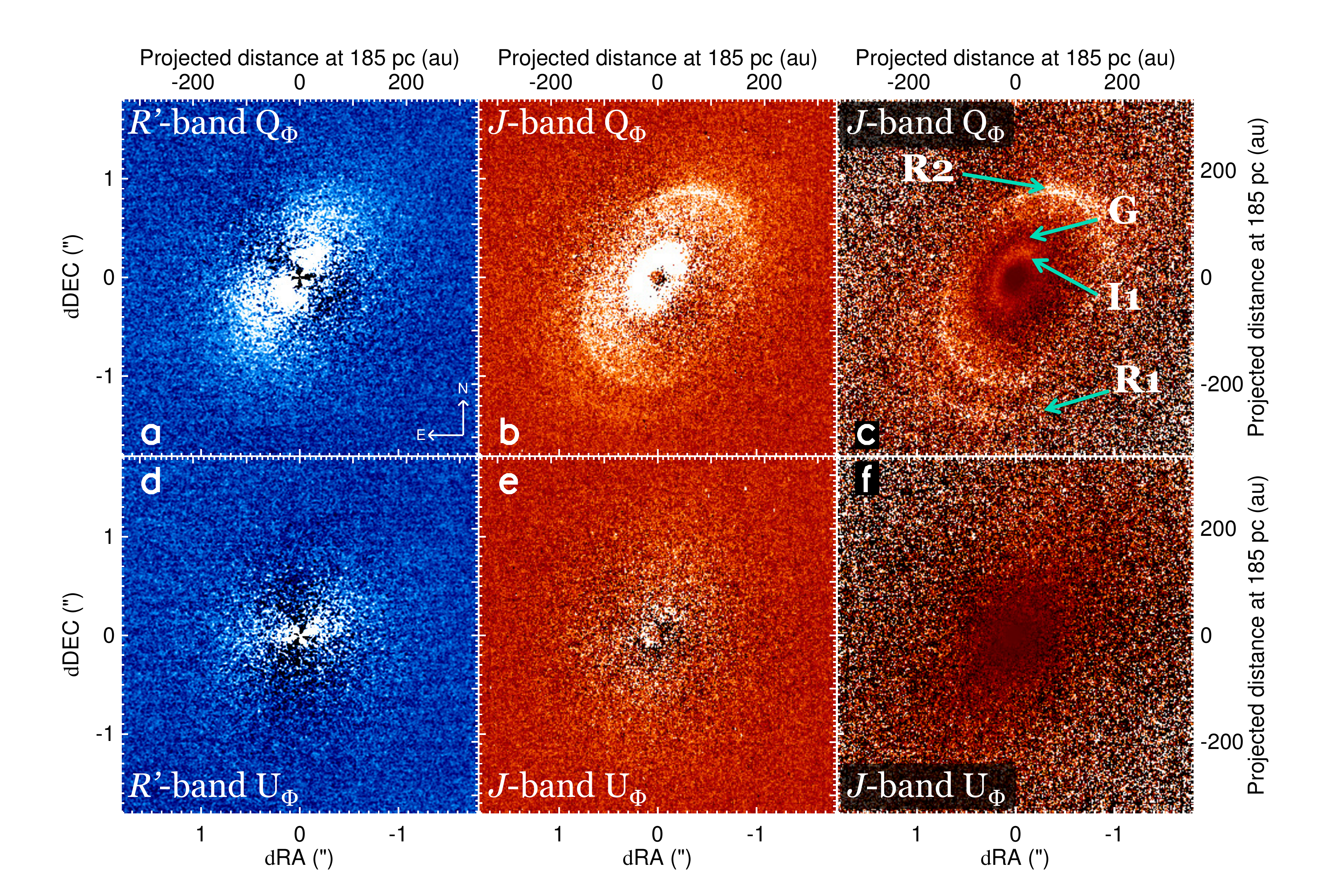}
   \caption{
	\textbf{a:} ZIMPOL \textit{R'}-band Q$_\phi$ image, smoothed with a boxcar of 3 pixels. 
	No radial scaling applied
	\textbf{b:} IRDIS/DPI \textit{J}-band Q$_\phi$ image, without radial scaling.
	\textbf{c:} IRDIS/DPI \textit{J}-band Q$_\phi$ image, with inclination ($i = 45^\circ$) corrected $r^2$ scaling. 
	This panel has the following disk features annotated: The southern ansa of a ring ($R1$), a full ring $R2$, 
	a clear depression/gap in surface density ($G$) and an inner disk structure $I1$.
	Panels \textbf{d},\textbf{e}, and \textbf{f} display the \up images corresponding to and with the same dynamic range as the \qp images of panels a, b, and c respectively.
      \label{fig:square}}
    \end{figure*}

The Polarimetric Di\ff erential Imaging \citep[PDI,][]{Kuhn:2001ApJ} reduction of the ZIMPOL data is performed according to the description of de Boer et al. (submitted) which is brie\fl y summarized below.
After dark subtraction and \fl at-\fix elding, we separate the orthogonal polarization states from each frame, and bin the resulting images to a pixel scale of $15 \times 15$\,mas.
For the \fix rst \thwp $= 0^\circ$ , we obtain the intensity image $I_{Q^+}$ by adding the two polarization states per frame, and sum over the two frames 
\citep[which measure the `$0$' and `$\pi$' phase,][]{2012SPIE.8446E..8YS}.
We compute $Q^+$ by subtracting the two polarization states per frame and subtracting the residual for the $\pi$-phase (2nd frame) from the residual image of the $0$-phase (1st frame).
In the same manner, we measure 
$I_{Q^-}$ and $Q^-$ with \thwp $= 45^\circ$;  
$I_{U^+}$ and $U^+$ with \thwp $= 22.5^\circ$; and
$I_{U^-}$ and $U^-$ with \thwp $= 67.5^\circ$.

We center each Stokes I image by determining the o\ff set between the image and a Mo\ff at function using a cross-correlation.
We use this o\ff set to shift the image to the center and
apply the same shift to the corresponding $Q^{+/-}$ and $U^{+/-}$ images, and compute for both detectors (d1/d2):
\begin{eqnarray}
	Q_\mathrm{d1/d2} &=&  (Q^+ - Q^-)/2, 	\label{eq:qddif} \\	
	U_\mathrm{d1/d2} &=&  (U^+ - U^-)/2.	 
	\label{eq:uddif}	
\end{eqnarray}

Because we used the same ($R'$) \fix lter for both ZIMPOL detectors, we get the \fix nal $Q$, $U$ and $I_{Q/U}$ images according to: 
\begin{eqnarray}
	Q &=&  (Q_\mathrm{d1} - Q_\mathrm{d2})/2, \\
	U &=&  (U_\mathrm{d1} - U_\mathrm{d2})/2,	
\end{eqnarray}
where the minus sign is necessary because the polarizing beamsplitter yields orthogonal polarization signals for the two detectors.
We correct the $Q$ and $U$ images for remaining instrumental and background polarization according to \citet{2011A&A...531A.102C}.

Finally, we compute the azimuthal Stokes components 
\citep[cf.][where the azimuth $\phi$ is de\fix ned with respect to the star-center]{Schmid:2006A&A}:
\begin{eqnarray}
	Q_\phi &=& Q \times \cos{2\phi} + U \times \sin{2\phi}, 
		\label{eq:qphi}\\
	U_\phi &=& Q \times \sin{2\phi} - U \times \cos{2\phi}.
	\label{eq:uphi}
\end{eqnarray}
For cases of single scattering or simple symmetries, \qp should contain all the signal, 
while \up provides an indication of the measurement error.
The \qp and \up images are smoothed with a boxcar of three pixel width and shown in Figures~\ref{fig:square}a \& \ref{fig:square}d, respectively.

	\begin{figure*}
	\centering
	 	\includegraphics[angle = 0, width=1\textwidth, trim = 10 10 30 10]{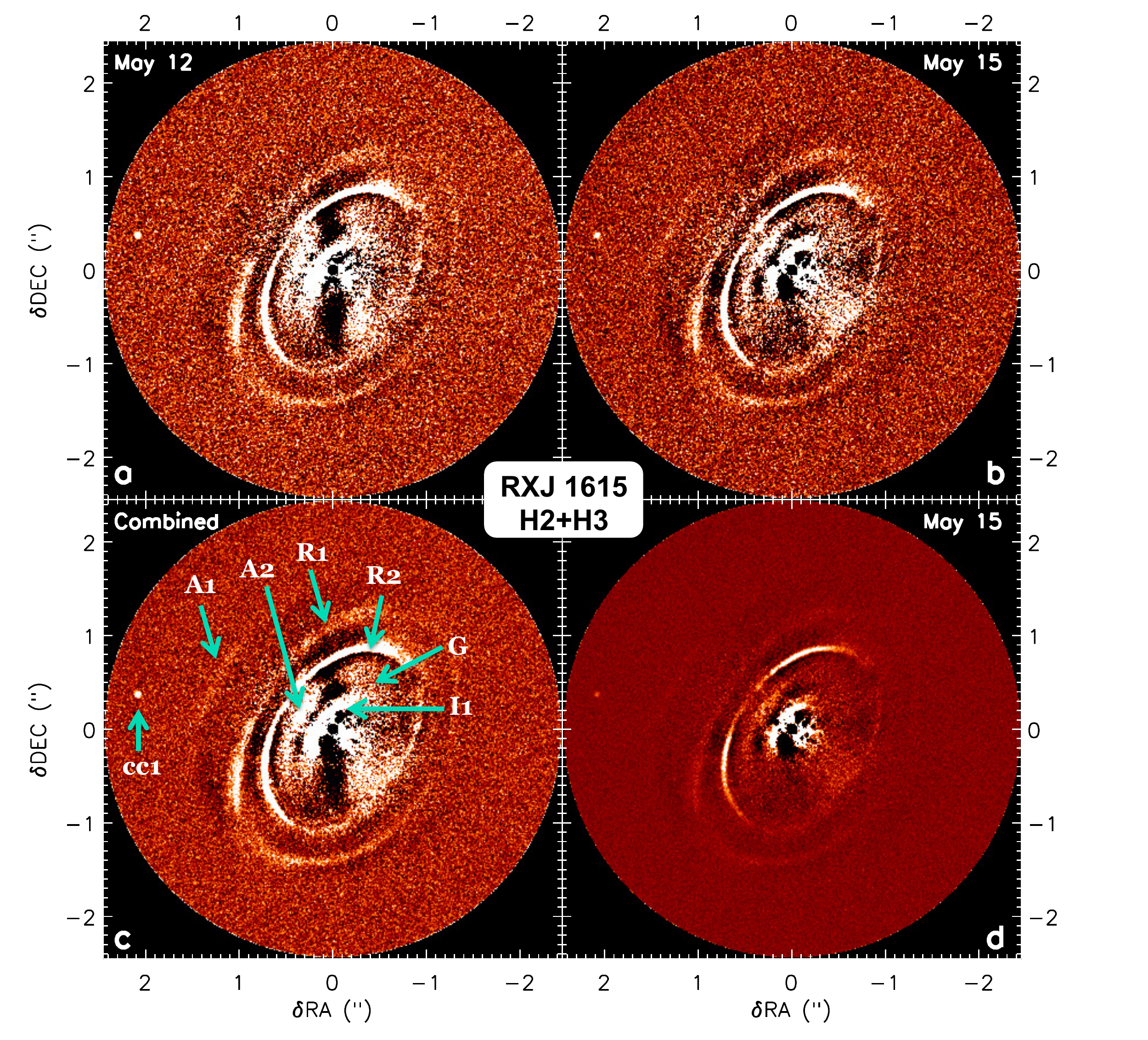}
	   	\caption{TLOCI reductions of SPHERE-IRDIS ADI H2H3 (mean combination of the H2 and H3 \fix lters) data.
	   	\textbf{a}: May 12 2015; 
	   	\textbf{b}: May 15 2015;
	   	\textbf{c}: Mean combination of May 12 and May 15, with features annotated. From the outside in, we see a point source, which we consider as a companion candidate (cc1), an arc ($A1$), two full rings ($R1$ \& $R2$), another arc ($A2$), the location of the gap seen in Figure~\ref{fig:square}c ($G$), and an innermost disk structure (I1);
	   	\textbf{d}: Same as b, for a $10 \times$ larger dynamic range.
      		\label{fig:irdisifs}}
    \end{figure*}

\begin{figure*}
	\centering
	 	\includegraphics[angle = 0, width=0.9\textwidth, trim = 70 0 70 0]{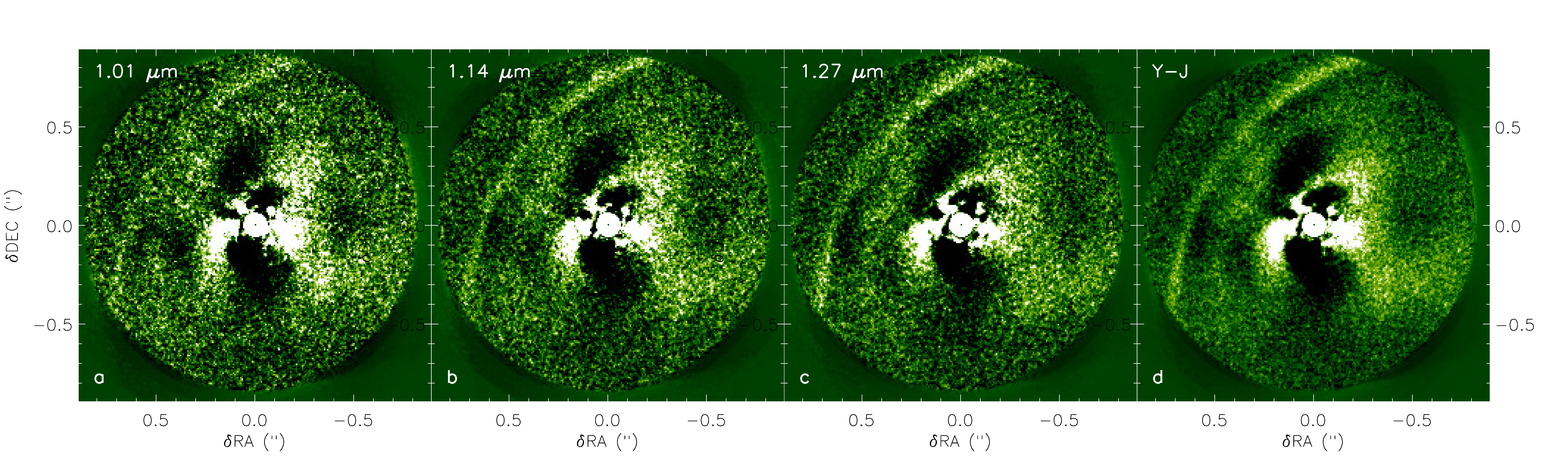}
	 	\caption{TLOCI reductions of SPHERE-IFS ADI Y-J data (0.96-1.07\,$\mu$m)
	 	\textbf{a:} Median of 13 channels (0.96-1.07\,$\mu$m).
	 	\textbf{b:} Median of 13 channels (1.08-1.21\,$\mu$m).
	 	\textbf{c:} Median of 13 channels (1.22-1.33\,$\mu$m).
	 	\textbf{c:} The median over the entire Y-J range of the IFS.
	 	\label{fig:ifs}}
    \end{figure*}

		\begin{figure*}
	\centering
	 	\includegraphics[width=1\textwidth, trim = 10 0 30 0]{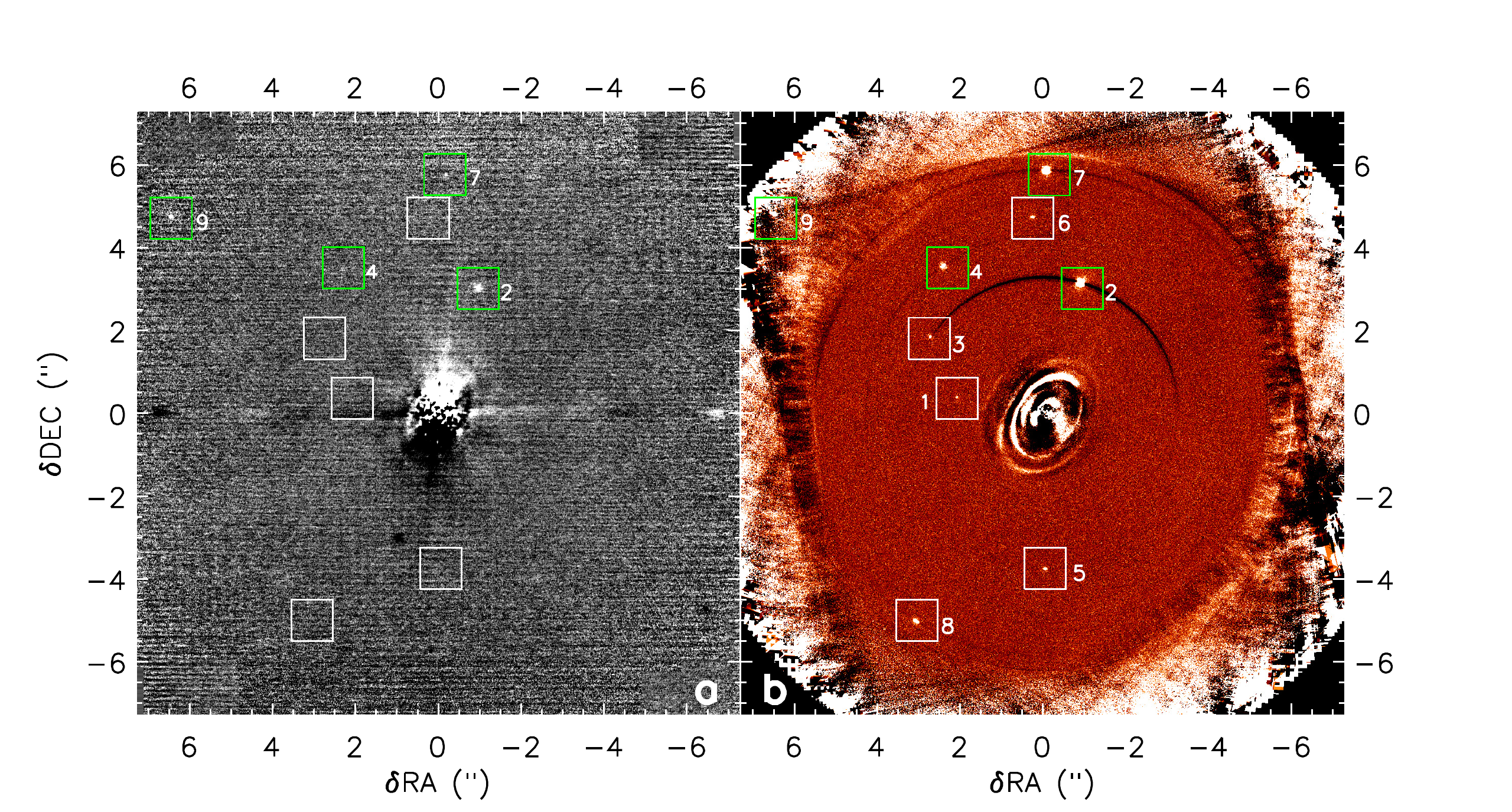}
	   	\caption{Companion candidates detected in NACO and SPHERE data. \textbf{a}: NACO image in \textit{K$_s$}-band of May 7, 2010.
	   	We self-subtracted the image after it was rotated by $180^\circ$.
	   	A small section of ${R2}$ is seen west of the star.
	   	\textbf{b}: SPHERE-IRDIS cADI in \textit{H23}-band of May 15, 2015. 
	   		The green squares in a and b show the 4 point sources \fix rst detected with NACO; 
	   		the white squares highlight the point sources detected with SPHERE-IRDIS only.
      		\label{fig:nacoirdis}}
    \end{figure*}

\subsection{IRDIS/DPI observations in $J$-band}

The IRDIS DPI mode splits the beam before two orthogonal polarizers create the two orthogonal
images simultaneously on di\ff erent halves of the detector: the ordinary intensity image $I_{o} = 0.5 (I + Q)$ 
and the extraordinary intensity image $I_{e} = 0.5 (I - Q)$. 
After we subtract the dark, and \fl at-\fix eld the images, we center the image of both detector halves on 
a Mo\ff at function, using a cross-correlation.
This \fix rst centering does not guarantee that the star is placed exactly at the center yet, 
but does ensure that all images of the star are at the same place with respect to the center, 
which su\ffi ces to subtract the two orthogonal polarization states on each frame.
For the first \thwp $Q^+$ is computed by subtracting the two simultaneouly measured beams:
\begin{equation}
	Q^+ = (I_o - I_e)|_{\theta_\mathrm{hwp} = 0}.	
\end{equation}
By performing this operation for each of the three frames observed per \thwp we obtain three 
$Q^+$; $Q^-$;$U^+$; and $U^-$ images
for \thwp $= 0^\circ; 45^\circ; 22.5^\circ;$ and $67.5^\circ$, respectively.

We stack the three di\ff erence images obtained per $\theta_\mathrm{hwp}$ 
and apply for each of the six HWP cycles the double di\ff erence:
\begin{eqnarray}
	Q &=& (Q^+ - Q^-)/2, \\
	U &=& (U^+ - U^-)/2.
\end{eqnarray}
For all HWP cycles we apply a correction on the $Q$ and $U$ images to remove a detector artefact 
which creates continuous vertical bands on the IRDIS detector which vary with time.
This correction is similar to the correction of \citet{2014ApJ...781...87A} for comparable artefacts on the NACO detector:
for each pixel column we take the median over the top 20 and bottom 20 pixels
and subtract this signal from the entire column.
Next, we compute \qp and \up according to Equations~\ref{eq:qphi} and \ref{eq:uphi}.

An additional centering per HWP cycle uses a minimization of the signal in the \up image,
which is based on the assumption that no astrophysical signal is measured in the \up image, ony noise (including reduction artefacts).
We shift the $Q$ and $U$ image at subpixel steps in the x and y direction and 
compute \up for each step. 
The shift which has the \up image with the lowest value over a centered but co-shifted annulus ($10 \le r \le 20$\,pixels) is assumed to place the star closest to the center of the image.

After we stack the centered $Q$ and $U$ images, 
we correct for instrumental and sky polarization, as we did for the ZIMPOL reduction.
From these stacked images, we compute the \fix nal \qp and \up images for the $J$-band, 
which are shown at the same intensity scale in Figures~\ref{fig:square}b and \ref{fig:square}e, respectively.
Figures~\ref{fig:square}c and \ref{fig:square}d show these \qp and \up images after scaling with the inclination-corrected radius squared,
for which we used $i = 45$ from \citet{Marel:2015A&A}.

\subsection{IRDIS ($H2H3$) and IFS ($YJ$) pupil tracking observations}
The IRDIS and IFS data are reduced 
using the SPHERE Data Center. 
We use the SPHERE pipeline \citep{Pavlov:2008SPIE} to process cosmetic reductions including sky subtraction, bad pixel removal, \fl at \fix eld and distortion corrections,
IFS wavelength calibration, IFU \fl at correction, instrument anamorphism correction \citep[0.60\,$\pm$\,0.02\%,][]{Maire2016}, and frame registering.

Then, several types of Angular Di\ff erential Imaging \citep[ADI,][]{Marois:2006ApJ} based algorithms are implemented in a dedicated tool (SpeCal, Galicher et al., in prep.) 
to perform starlight subtraction independently for each of the two IRDIS \fix lters, and the 39 IFS spectral chanels. 

For IRDIS, we present the results obtained with classical ADI (cADI) for the full \fix eld of view 
(Figure~\ref{fig:nacoirdis}b) and the
Template Locally Optimized Combination of Images algorithm \citep[TLOCI,][]{Marois:2014SPIE} for a 4$''$x4$''$ \fix eld (Figure~\ref{fig:irdisifs}).
The May 12 data shows a strong vertical negative residual in the reduction, for both IRDIS (Figure~\ref{fig:irdisifs}a) and IFS. 
We ascribe this to an aberration of the PSF, due to a loss of AO performance when coping with high altitude winds. 
The May 15 data is not plagued by this e\ff ect.
Figure~\ref{fig:ifs} shows the IFS TLOCI+ADI reductions for May 15, where the \fix rst three panels show the median 
combination of 13 spectral chanels between
0.96-1.07\,$\mu$m, 1.08-1.21\,$\mu$m, and 1.22-1.33\,$\mu$m, respectively. The righthand panel shows the median combination of all 39 spectral bands.

\subsection{NACO $K_s$ jitter imaging observations}
The NACO data reduction is similar for all epochs. 
First, we average all frames in each data cube, and use the \textit{jitter} routine in the ESO Eclipse package (\citealt{2001ASPC..238..525D}) to \fl at-\fix eld, shift and combine all averaged images.
An initial o\ff set of each frame is determined from the image header and is then re\fix ned using \textit{jitter's} cross correlation routine. 
Since the star was moved to di\ff erent positions on the detector, each pair of consecutive frames can be used to estimate the sky background in $K_s$-band and subtract it.
To remove the stellar halo, we stack the centered image, and subtract the same image after we have rotated it with $180^\circ$.
We present the resulting $180^\circ$ di\ff erential image in Figure~\ref{fig:nacoirdis}a.

The astrometric calibration of the NACO epochs was taken from \cite{2014MNRAS.444.2280G}. 
They imaged the core of the globular cluster 47\,Tuc for this purpose in the same \fix lter and imaging mode. 
These calibrations are within two days of the \rxj observations for the 2010 and the 2012 epoch. 
Due to bad conditions during the 2011 observations, no companion candidates were recovered. 

\subsection{Keck NIRC2 SAM $K'$ observations}
Data were reduced using the aperture masking pipeline developed at the University of Sydney. An in-depth description of the reduction process can be found in \citet{Tuthill:2000PASP} and \citet{Kraus:2008APJ}, but a brief summary follows: data were dark subtracted, 
\fl at-\fix elded, cleaned of bad pixels and cosmic rays, then windowed with a super-Gaussian function. The complex visibilities were extracted from the cleaned cubes and turned into closure phases. The closure phases were then calibrated by subtracting a weighted average of those measured on several point-source calibrator stars observed during the same night.

\section{Results}
\label{sec:results}

\subsection{Disk}
\label{sec:resdisk}

	The disk of \rxj is detected in all the datasets included in this study. 
	For the ADI as well as the PDI images, all features described below are brighter on the northeastern side from the major axis.

\textbf{- Outer rings:}
	In the $J$-band $r^2$ scaled Q$_\phi$ image (Figure~\ref{fig:square}c) and the (not $r^2$ scaled) 
	$H2H3$ TLOCI ADI images of the disk, 
	we see multiple arc-like and continuous ellipses 
	which we consider to be ring-like features 
	in the scattering ($\tau = 1$) surface of the disk, 
	projected with the inclination of the system.
	For decreasing separation, Figure~\ref{fig:irdisifs}c shows: 
	\begin{itemize}
		\item an arc ({\boldmath$A1$}); 
		\item two full rings {\boldmath$R1$}  and {\boldmath$R2$};
		\item a second arc ({\boldmath$A2$}).
	\end{itemize}
	Where the two arcs cross the minor axis of the disk, they seem to lie parallel to both rings. 
	We therefore consider it most likely that both arcs are segments of full rings.
	The \aii feature does not appear to be as clearly separated from \rii in the $J$-band 
	$r^2$ scaled polarization image, and is therefore not annotated in the image.
	However, in this \qp image the disk signal at radii $r \sim r(A2)$ appears continuous in the azimuthal direction, 
	which con\fix rms that \aii is indeed a full ring.
	The $R2$ disk feature is detected (at least as a ring segment) in all datasets except the ZIMPOL $R'$-band polarization image. 
	In Figure~\ref{fig:ifs}, we see that the \rii segment which lies within the IFS \fix eld of view is detected more clearly at longer wavelengths.
	Only the southern ansa of \ri is detected in the $J$-band Q$_\phi$ image.
	In Figures~\ref{fig:square}b,c and \ref{fig:irdisifs} we can discern that \ri and \rii 
	are clearly not concentric with the inner disk component \ii and the star-center.

\textbf{- Gap:}
	A gap (feature {\boldmath$G$}) 
	in between features \aii and \ii is detected, most clearly in Figure~\ref{fig:square}c. 
	Figure~\ref{fig:irdisifs}c shows the gap as well, but not for all azimuth angles.

\textbf{- Inner disk:}
	Closest to the star, we see an elliptical inner disk component (feature {\boldmath$I1$} in Figure~\ref{fig:irdisifs}b). 
	The surface brightness is continuous inward in the Q$_\phi$ images 
	in both $R'$ and $J$-band (Figures~\ref{fig:square}a \& \ref{fig:square}c).
	However, for a disk with continuous surface density with a linearly increasing scattering surface 
	we would expect the surface brightness to drop o\ffs with the distance to the star squared.
	Conversely, if we create an image with the surface brightness scaled with $r^2$ (corrected for the inclination), the previous
	example of a continuous surface density would show a continuous surface brightness. 	
	However, \ii in our inclination-corrected $r^2$ $J$-band image (Figure~\ref{fig:square}c) appears more ring-like, with a cavity inside, which 
	agrees 
	with the outer radius of the inner dust cavity as determined from the 880 $\mu$m interferometric observations of
	\citet[$r_\mathrm{cav} = 30$\,au]{Andrews:2011ApJ}. 
	Still, we consider the detection of the cavity as tentative because $r_\mathrm{cav}$ is bordering the coronagraph.
	We also see a non-continuous surface brightness of \ii in the ADI reductions of Figure~\ref{fig:irdisifs}.
	However, observations of a disk with continuous surface brightess is likely to be plagued by self-subtraction, 
	which is often seen in ADI for low 
	and intermediate inclination circumstellar disks \citep{Milli:2012}. 
	This kind of self-subtraction does not occur in PDI, which is extremely e\ffi cient at isolating the polarized disk signal, hence revealing the disk structure with high fidelity.
	Note that based on our results we cannot rule out a disk component at a separation $r < 30$\,au (inside \iib).
	However, \citet{Andrews:2011ApJ} mention that their model requires a very low density inside $r_\mathrm{cav} = 30$\,au
	to adequately fit the SED.

\subsubsection*{Ellipse \fix tting}
\label{sec:ellipse}
 

 \begin{table}[!h]
	\resizebox{0.94\textwidth}{!}{\begin{minipage}{\textwidth}   	
	\begin{tabular}{ c|l|l|l|l|l|}
	   & Parameter            & ADI-$H23$ & $\sigma_{H23}$ &PDI-$J$ & $\sigma_{J}$ \\
\hline \hline
	\ri         & Semi-major axis ($''$)      &   1.50  &0.01& 1.50 & 0.01 \\
	             & Semi-minor axis ($''$)      &   1.01  & 0.01& 1.02  & 0.01\\
	             & RA o\ff set $u_\mathrm{x}$ ($''$) &  -0.15   & 0.01& -0.15& 0.01\\
	             & Dec o\ff set $u_\mathrm{y}$ ($''$) & -0.10  & 0.01& -0.10& 0.01\\
	             & O\ff set angle ($^\circ$)    &   238  & 1 &  236 & 1\\
	             & \textit{PA} ($^\circ$)         & 145.7  & 1.0 & 144.2& 0.8\\
	             & Inclination angle ($^\circ$) &47.3 & 1.0& 47.0& 0.8\\
	             & $H_{\tau = 1}$ (au)                    &   44.9     & 2.2& 44.7 & 1.7\\
	             & $H_{\tau = 1} / r$                      & 0.162      & 0.009 & 0.162 & 0.007\\
	\cline{2-6}
	\rii          & Semi-major axis ($''$)&        1.06   & 0.01& 1.06 & 0.01\\
	             & Semi-minor axis ($''$)       &   0.70  &0.01& 0.72  & 0.01\\
	             & RA o\ff set $u_\mathrm{x}$ ($''$)  &  -0.10  & 0.01& -0.10& 0.01\\
	             & Dec o\ff set $u_\mathrm{y}$ ($''$) & -0.07 & 0.01& -0.06& 0.01\\
	             & O\ff set angle ($^\circ$)                & 235  & 1 &  236& 2\\
	             & \textit{PA} ($^\circ$)                      & 145.4& 1.3& 144.3& 1.4\\
	             & Inclination angle ($^\circ$)            & 48.5 & 1.3 & 46.8& 1.4\\
	             & $H_{\tau = 1}$ (au)                        & 30.9     & 2.4 &  29.6 & 2.2\\
            	     & $H_{\tau = 1} / r$                      & 0.158      & 0.014 & 0.152 & 0.013\\
	\cline{2-6}
	\ii         & Semi-major axis ($''$)     &  0.30    & 0.01& 0.35 & 0.01\\
	             & Semi-minor axis ($''$)     &  0.20   & 0.01 & 0.24  & 0.01\\
	             & RA o\ff set $u_\mathrm{x}$ ($''$)  & -0.01& 0.01 & 0.00& 0.01\\
	             & Dec o\ff set $u_\mathrm{y}$ ($''$) & 0.00 & 0.01 & 0.00& 0.01\\
	             & O\ff set angle$^*$ ($^\circ$)     & 261&     &  209 &\\
	             & \textit{PA} ($^\circ$)  & 145.5 & 4.2& 144.5&4.3\\
	             & Inclination angle ($^\circ$) & 49.0 & 3.9& 47.7& 4.1\\
	             & $H_{\tau = 1}^*$ (au)                    &  3.5    && 1.4 & \\
	\hline
	\end{tabular}
	\vspace{1mm}
	\end{minipage}}
		\caption{Ellipse parameters for the \fix ts to the features listed in Figure~\ref{fig:ellipses}a and \ref{fig:ellipses}b.
		$^*$) For \ii, the errors on the ellipse o\ff set ($u$) are larger than the measured value. 
		We therefore do not consider the value for the ellipse o\ff set angle and $H_{\tau = 1}$ to be signi\fix cant.
			}
		\label{tab:param}
\end{table}
	\begin{figure*}	
		\centering
		\includegraphics[width=0.8\textwidth, trim = 80 0 80 0]{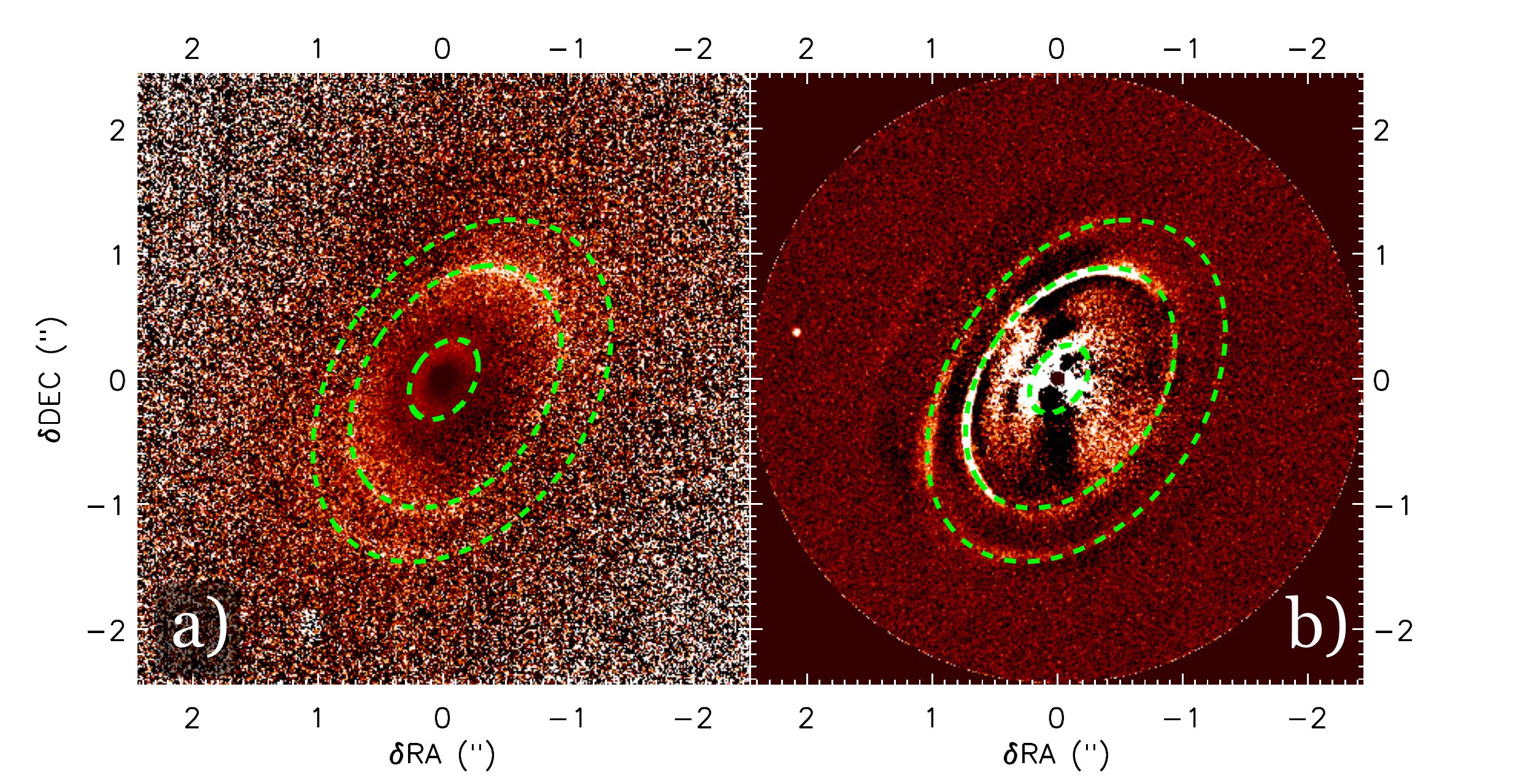}
		\caption{
		\textbf{a:} IRDIS $J$-band PDI image after inclination-corrected $r^2$ scaling. 
		\textbf{b:} IRDIS $H23$-band TLOCI + ADI image.
			The overplotted green ellipses are the \fix ts to \rib, \rii and \iib, as listed in Table~\ref{tab:param}. 
     		\label{fig:ellipses}}
    	\end{figure*}

		\begin{figure*}
		\centering
		\includegraphics[width=0.75\textwidth, trim = 10 0 80 10]{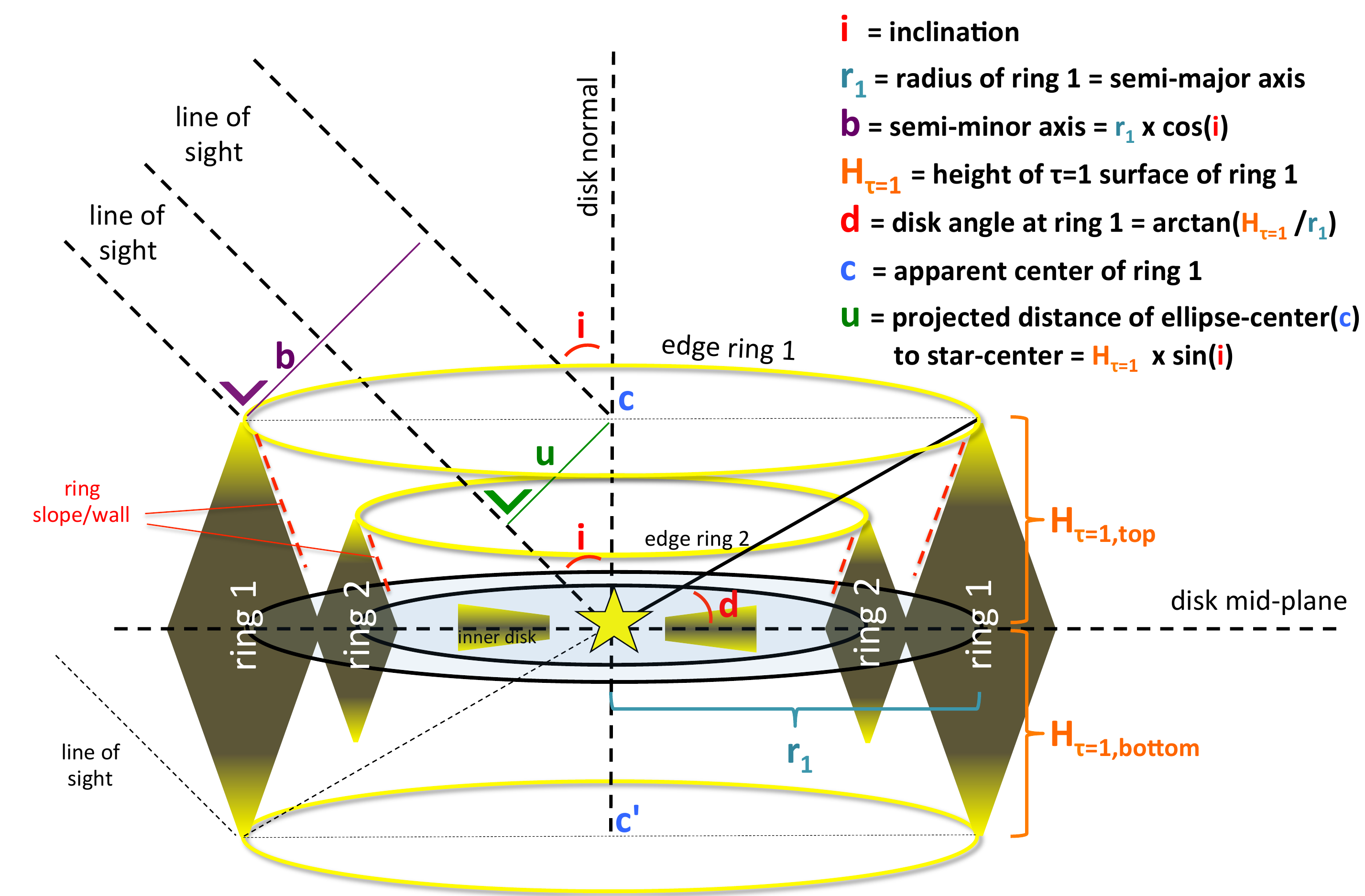}
		\caption{Schematic view of a double ringed thick disk. Observed at inclination $i$, circular rings will appear elliptical. 		
		The height of the scattering surface of Ring 1 
		($H_{\tau = 1}$(\rib)) can be found by determining the semi-minor axis ($b$), $i$ 
		and the distance between the star and the center of the ellipse ($u$).
		Note that this figure has the purpose of explaining how to determine $H_{\tau = 1}$ for the di\ff erent rings, not to portray ideas about the radial density distribution.
     		\label{fig:flaring}}
    	\end{figure*}

	For both the $J$-band PDI image and the $H23$-band TLOCI ADI image we have \fix tted ellipses\footnote{ 
	using the routine \textit{mpfitellipse.pro} with the \textit{Interactive Data Language} (IDL)}
	 to the two rings and to the inner disk.
	The resulting ellipses are overplotted in green in Figures~\ref{fig:ellipses}a and \ref{fig:ellipses}b, 
	and the ellipse parameters are listed in Table~\ref{tab:param}.
	From the assumption that the ellipses trace the highest surface above the midplane for circular rings 
	(i.e. not the slope facing the star/wall), 
	we determine the inclination of each ellipse according to $\cos{i} = $ semi-minor axis$/$semi-major axis.
	With a weighted mean we \fix nd an inclination of the disk $i = 47.3 \pm 0.4^\circ$, 
	where the error represents the random error on the \fix tted ellipses but does not include the systematic errors from our method. 
	A conservative estimate of the systematic errors brings us to a \fix nal value of $i = 47 \pm 2^\circ$,
	which is in good agreement with
	the inclination derived by \citet{Marel:2015A&A}. 
	The weighted mean for the position angles is \textit{PA} $= 145.8 \pm 0.5^\circ$. 
	Including our estimate of the systematic errors gives a \fix nal \textit{PA} $= 146 \pm 2^\circ$, in between the 
	values of \citet[\textit{PA} $= 143^\circ$]{Andrews:2011ApJ} and \citet[\textit{PA} $= 153^\circ$]{Marel:2015A&A}.
	
	We also \fix nd that \ri and \rii are not centered around the star (listed as X ($u_\mathrm{x}$) and 
	Y ($u_\mathrm{y}$) o\ff set in Table~\ref{tab:param}).
	The o\ff set of the ellipse-centers with respect to the position of the stars means that the rings are either: 
	\begin{itemize}
	\item eccentric rings; 
	\item concentric circular disk components with considerable radially increasing thickness,
	as described by the height of the $\tau = 1$ surface 
	($H_{\tau = 1}(r)$) above the disk midplane; 
	\item an intermediate of the two extremes: an eccentric and a thick ring.
	\end{itemize}	
	The directions in which \ri and \rii are o\ff set from the star (listed as `O\ff set angle' in Table~\ref{tab:param})
	are roughly \textit{PA} $ + 90$, i.e. along the minor axes of the ellipses.
	These apparent ellipse displacements along the minor axes 
	is a necessary condition for the o\ff sets to be caused by a projection of the $\tau = 1$ surface on the midplane due to 
	the inclination of the system.
	Below, we explore the scenario of thick circular rings viewed at an inclination $i$ away from face-on.
	In this scenario, the disk in the direction opposite to the ellipse o\ff sets (i.e. the eastern side, \textit{PA} $\sim 56^\circ$) forms the near side.
	Without trying to explain the surface density distribution of the disk, 
	in Figure~\ref{fig:flaring} we portray such a disk with thick rings.
	From the sketch we can derive that for any given ellipse with its center 
	o\ff set from the star-center with distance $u = ({u_\mathrm{x}^2 + u_\mathrm{y}^2})^{1/2}$,
	we can determine the height of the scattering surface ($H_{\tau = 1}(r)$) of this ring according to:
	\begin{eqnarray}
	\frac{u}{b} &=& \frac{H_{\tau = 1}(r)}{r} \times \frac{\sin{i}}{\cos{i}} = \tan{d} \times \tan{i},
	 \label{eqn:ub}\\
	H_{\tau = 1}(r) &=& \frac{u}{b } \times \frac{r}{\tan{i}} = \frac{u}{\sin{i}}. \label{eqn:flar}
	\end{eqnarray}

	In a similar fashion, \citet{Lagage:2006} have determined the tickness of 
	HD\,97048 based on PAH emission maps. 
	The main di\ff erence between the method used by \citet{Lagage:2006} and our method is that
	the former have used isophotes, while we can use the sharp rings in the disk of \rxjb.
	It is clearly visible in Figure\,\ref{fig:ellipses} that
	the surface brightness of the rings is strongly varying with azimuth angle. 
	To accurately determine $H_{\tau = 1}$ based on isophotes it is necessary to correct for any 
	azimuthal surface brightness variation, which requires radiative transfer modeling.
	To the best of our knowledge, this is the first example where $H_{\tau = 1}$ can be determined 
	strictly from geometrical constraints (i.e. model independent) in scattered light images.

\subsubsection*{Polarized intensity pro\fix le}	
		\begin{figure}
		\centering
		\includegraphics[width=0.5\textwidth, trim = 20 0  0 -5]{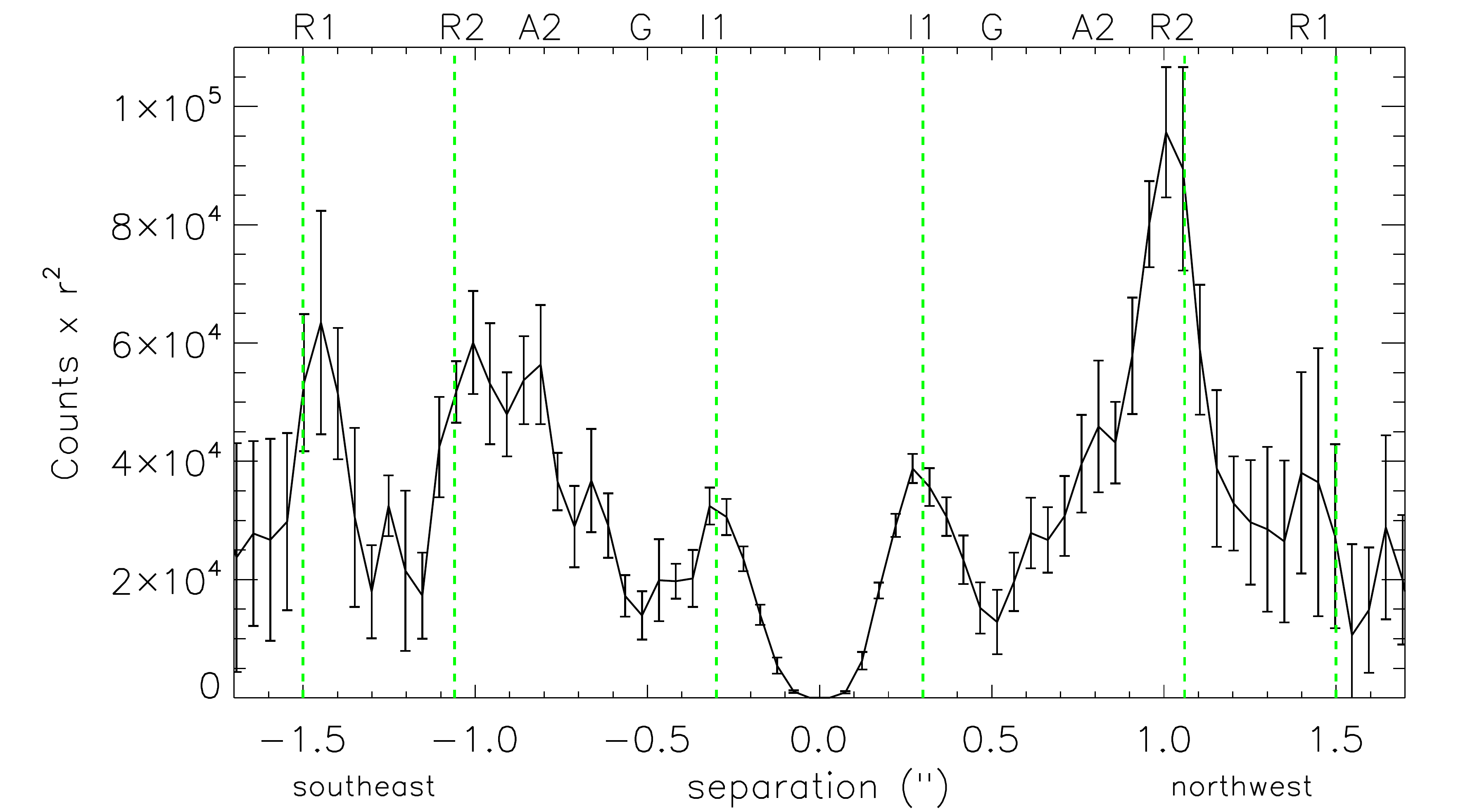}
		\caption{\qp intensity pro\fix le of $r^2$ scaled $J$-band image (Figure~\ref{fig:square}c) along \textit{PA} $= 146.5^\circ$.
		The green dashed lines mark the semi-major axes of the ellipses \rib, \rii and \iib. 
		Note that although this pro\fix le is measured for a \textit{PA} similar to those of the ellipses, 
		the pro\fix le does include the star center, and therefore does not lie on the major axes of \ri and \riib.
		This explains why the peaks of the rings lie at closer separations than their semi-major axes.
     		\label{fig:cut}}
    	\end{figure}

	Figure~\ref{fig:cut} shows the intensity pro\fix le of the $r^2$ scaled \qp image in $J$-band (Figure~\ref{fig:square}c), 
	after smoothing with 4 pixels. 
	The pro\fix le is measured along \textit{PA} $= 146.5^\circ$, centered on the star (i.e. parallel to but not on the major axes of \ri and \riib). 
	The errors are the standard deviation measured over the same aperture
	in the \up image (Figure~\ref{fig:square}f), divided by the square root of the number of pixels.
	The disk features which can be distinguished in the pro\fix le are listed along the top axis of the plot.
	Due to the ellipse-center o\ff set with respect to the star, the pro\fix le does not cross the ansae of the ellipses, which becomes visible when we compare the position of peaks in the intensity pro\fix le with the semi-major axis of the ellipses for \ri and \riib, annotated with green dashed lines (values adopted from Table~\ref{tab:param}).
	Although the feature is not clearly visible in Figure~\ref{fig:square}c, we can recognize the two peaks of \aii in the the \qp intensity pro\fix le.
	The gap $G$ is visible at $r \sim 0.5'' \approx 93$\,au. 
	\citet{Marel:2015A&A} \fix nd a tentative dust gap between 110 - 130\,au (0.6$''$ - 0.7$''$) in the 690\,GHz continuum profile. This sub-mm gap lies outside the gap we detect in the $J$-band data.
	
\subsection{Point sources}
\label{sec:resccs}

	\begin{figure*}
   		\centering
		\includegraphics[width=1.\textwidth]{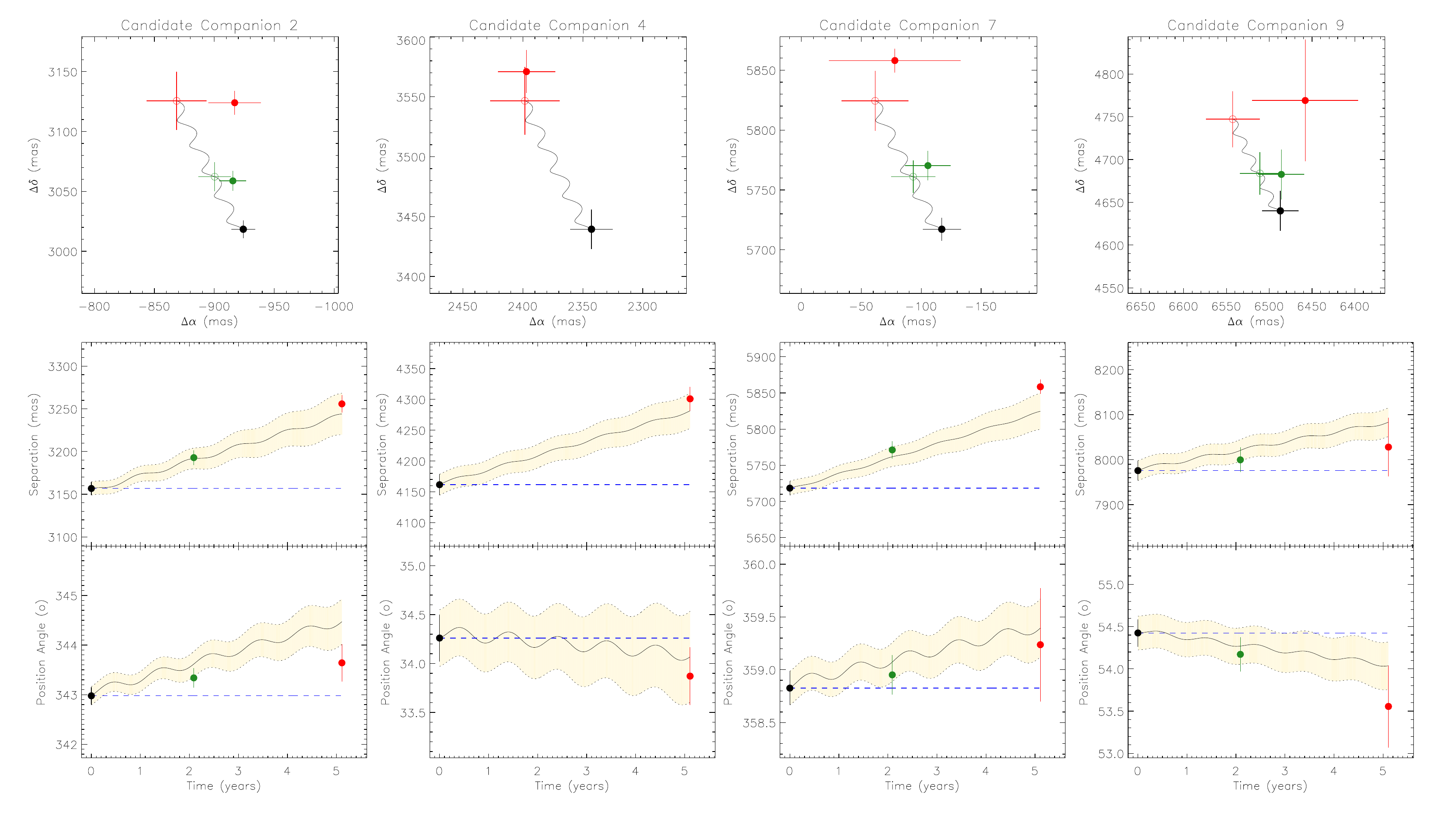}
   		   	\caption{
   		   	Astrometry for the 4 point sources detected (solid points) by both NACO (black and green) and IRDIS (red) and the expected values for background objects (open points). 
   		   	Data points are color coded by date, not by instrument. 
   		   	These point sources are marked with the green squares in Figure~\ref{fig:nacoirdis}.
   		   	The numbers of the candidate companions listed in the title of each panel column are the same as shown in Figure~\ref{fig:nacoirdis}.
   		   	All 4 point sources are shown to not be co-moving with \rxjb.
      		\label{fig:astro}}
      	\end{figure*}

	Four point sources have been detected in the NACO $K_s$-band image, and marked with green boxes 
	in Figure~\ref{fig:nacoirdis}a \& \ref{fig:nacoirdis}b.
	In the IRDIS data (Figure~\ref{fig:nacoirdis}b), we detect no point sources within the ringed disk structure, out to \aib.
	We do detect nine point sources outside the outermost disk component \aib, including all four point sources detected with NACO. 
	The \fix ve new companion candidates are marked with white boxes in Figure~\ref{fig:nacoirdis}b.
	For reference, we included the white boxes at the same locations in Figure~\ref{fig:nacoirdis}a, even though these point sources are not detected by NACO.
	The astrometry and photometry of the six innermost candidates detected with IRDIS were derived using the LAM-ADI pipeline \citep{vigan2012,vigan2016} 
	using injection of fake negative planets in the pre-processed ADI data cubes. 
	The position and \fl ux of the negative fake companion are adjusted using a Levenberg-Marquardt least-squares minimisation routine 
	where we try to minimise the residual noise after ADI processing in a circular aperture of radius $\lambda/D$ centered on the position of the companion candidate. 
	The error bars for the \fix tting process are then calculated by varying the position and contrast of the fake companion until the variation of the reduced $\chi^2$ reaches a level of 1$\sigma$. 
	For the speckle subtraction, we have used a PCA analysis \citep{soummer2012} where \fix ve modes are subtracted. 
	At the distances of the outer three candidates there is negligible speckle noise negating the need for such post-processing. For these we perform basic relative photometry after subtracting a radial pro\fix le and stacking the de-rotated frames. 
	The astrometry and photometry measurements for the companion candidates are reported in Table~\ref{tab:astrom}. 
	For IRDIS, we have adopted a true north direction of $-1.81 \pm 0.30^\circ$ and a plate scale of $12.27 \pm 0.016$\,mas/pix,
	based on astrometric calibrations performed on the theta Ori B field
	\citep[and Vigan private communication]{Maire2016}.
	For the NACO observations of 2010, we adopted a true-north of $0.67 \pm 0.13^\circ$ 
	and a plate scale of $13.231 \pm 0.020$\,mas; 
	for 2012 we used a true-north of $0.65 \pm 0.14^\circ$ and a plate scale of $13.234 \pm 0.021$\,mas.
	The \fix nal error bars for the photometry include the \fix tting error detailed above; 
	the uncertainties on the star center (1\,mas);
	 and the level of noise residuals estimated at the same separation as the detections.

\begin{table}[!h]
	\resizebox{0.775\textwidth}{!}{\begin{minipage}{\textwidth}   	
	\begin{tabular}{  c | c | c | c | c | c | c | }
	 	Obj 	& Date & Instr. & \fix lt. & $\Delta RA (mas)$ & $\Delta Dec (mas)$ & $\Delta mag$ \\ 
	\hline \hline
		cc1 	& 15/05/2015 & IRDIS & H2 & $2086 \pm 16$ & $405 \pm 25$ & $13.14 \pm 0.42$ \\ 
	 		& 15/05/2015 & IRDIS & H3 & $2090 \pm 16$ & $397 \pm 25$ & $13.17 \pm 0.34$ \\ 	
	 \cline{2-7}
		cc2 	& 05/04/2010 & NACO & Ks & $-924 \pm 10$ & $3018 \pm 8$ & $8.47 \pm 0.12$ \\ 
	 		& 07/05/2012 & NACO & Ks & $-916 \pm 11$ & $3059 \pm 8$ & $8.48  \pm 0.13$ \\ 
	 		& 15/05/2015 & IRDIS & H2 & $-907 \pm 22$ & $3242 \pm 9$ & $8.37 \pm 0.11$ \\ 
	 		& 15/05/2015 & IRDIS & H3 & $-909 \pm 22$ & $3243 \pm 10$ & $8.40 \pm 0.10$ \\ 
	 \cline{2-7}
		cc3 	& 15/05/2015 & IRDIS & H2 & $2722 \pm 27$ & $1854 \pm 33$ & $12.90 \pm 0.35$ \\ 
	 		& 15/05/2015 & IRDIS & H3 & $2722 \pm 27$ & $1859 \pm 33$ & $12.80 \pm 0.65$ \\ 
	\cline{2-7}
		cc4 	& 05/04/2010 & NACO & Ks  & $2343 \pm 18$ & $3439 \pm 17$ & $10.45 \pm 0.31$ \\
	 		& 15/05/2015 & IRDIS & H2 & $2397 \pm 24$ & $3571 \pm 18$ & $10.67 \pm 0.08$ \\ 
	 		& 15/05/2015 & IRDIS & H3 & $2397 \pm 24$ & $3571 \pm 18$ & $10.70 \pm 0.08$ \\
	\cline{2-7}
		cc5 	& 15/05/2015 & IRDIS & H2 & $-8 \pm 27$ & $-3745 \pm 5$ & $12.20 \pm 0.17$ \\ 
	 		& 15/05/2015 & IRDIS & H3 & $-8 \pm 27$ & $-3744 \pm 5$ & $12.19 \pm 0.17$ \\ 
	\cline{2-7}
		cc6 	& 15/05/2015 & IRDIS & H2 & $242 \pm 39$ & $4728 \pm 9$ & $12.30 \pm 0.22$ \\ 
	 		& 15/05/2015 & IRDIS & H3 & $234 \pm 39$ & $4730 \pm 9$ & $12.26 \pm 0.18$ \\ 
	\cline{2-7}
		cc7 	& 05/04/2010 & NACO & Ks & $-117 \pm 16$ & $5717 \pm 10$ & $9.88  \pm 0.24$ \\ 
	 		& 07/05/2012 & NACO & Ks & $-106 \pm 19$ & $5770 \pm 12$ & $9.96 \pm 0.27$ \\ 
	 		& 15/05/2015 & IRDIS & H2 & $-78 \pm 55$ & $5858 \pm 10$ & $9.74 \pm 0.39$ \\ 
	 		& 15/05/2015 & IRDIS & H3 & $-78 \pm 55$ & $5857 \pm 10$ & $9.71 \pm 0.38$ \\ 
	\cline{2-7}
		cc8 	& 15/05/2015 & IRDIS & H2 & $3065 \pm 46$ & $-5007 \pm 36$ & $11.45 \pm 0.43$ \\ 
	 		& 15/05/2015 & IRDIS & H3 & $3059 \pm 46$ & $-5010 \pm 36$ & $11.29 \pm 0.40$ \\ 
	\cline{2-7}
		cc9 	& 05/04/2010 & NACO & Ks & $6487 \pm 22$ & $4640 \pm 23$ &  $9.11 \pm 0.16$ \\ 
	 		& 07/05/2012 & NACO & Ks & $6486 \pm 27$ & $4683 \pm 29$ & $9.26 \pm 0.19$ \\ 
	 		& 15/05/2015 & IRDIS & H2 & $6458 \pm 62$ & $4769 \pm 71$ & $10.65 \pm 0.46$ \\ 
	 		& 15/05/2015 & IRDIS & H3 & $6439 \pm 87$ & $4790 \pm 95$ & $10.75 \pm 0.42$ \\ 
	\hline
	\end{tabular}	
	\vspace{1mm}
\end{minipage}}
	 \caption{Astrometry and photometry for nine companion candidates relative to \rxj }
		\label{tab:astrom}
\end{table}

In Figure~\ref{fig:astro} we compare the astrometry for the four point sources detected with NACO with that of our IRDIS detection. 
We determine them to not be co-moving, and therefore not associated to the \rxj system. 
The \fix ve remaining candidates, seen in Figure~\ref{fig:nacoirdis}b, require follow up observations to determine whether they are bound to \rxjb.

		No signi\fix cant point-source signals were found in any of the Keck NIRC2 SAM datasets, and so we place limits on their detectability by drawing 10,000 simulated closure phase datasets consistent with Gaussian random noise using the measured uncertainties. For each combination of separation, contrast and \textit{PA} on a 3D grid, the simulated datasets were compared to a binary model. The ($3.3\sigma$) detection limits were calculated as the point at which the binary model gave a worse \fix t to 99.9\% of the simulated datasets. The datasets probe separation ranges as small as 30\,mas, and achieve contrast limits that are approximately \fl at at larger separations. The 2014-06-10 data allow us to rule out objects with $\Delta K' <5.4$\,mag ( $>20$\,M$_\mathrm{jup}$ for a system age of 1.4\,Myr), while the 2012-07-08 and 2012-04-14 data reach similar contrasts of $5.0$\,mag and $5.2$\,mag, respectively.

\section{Discussion}
\label{sec:discuss}

\subsection{Disk geometry}
\label{sec:geo}

	The explanation for the apparent ellipse o\ff sets $u$
	by the projection of the $\tau = 1$ surface at height $H_{\tau = 1}$ above the midplane, as given in
	 Section~\ref{sec:resdisk}, implies that the northeast (\textit{PA} $ \sim 56$) is the near side of the disk
	 (i.e. pointing towards earth).
	\citet{Min:2012} and \citet{Dong:2016} show that the predominance of forward scattering 
	in the near side of disks gives it in general a larger surface brightness in total intensity than the far side.
	The ADI intensity image of Figure~\ref{fig:irdisifs} shows the northeastern sides of $R1$, $R2$ and $I1$ to be brighter thant their southwestern sides, con\fix rming that the northeast is the near side of the disk.

	In Section~\ref{sec:resdisk}, we \fix nd the disk inclination $i = 47 \pm 2^\circ$, 
	which is in good agreement with both the disk inclination derived by \citet{Marel:2015A&A} 
	and the inclination of the stellar rotation axis derived in Appendix~\ref{sec:stellar2} from $v \sin{i} = 13.0$\,km\,s$^{-1}$
	\citep[where the stellar rotational velocity $v$ depends on the determination of the stellar radius from its luminosity, 
which in turn depends on the interstellar extinction $A_V = 0$\,mag, 
as suggested by \citet{Manara:2014A&A}]{Wichmann99}. 
	Higher \snew extinction \enew \citep[e.g. $A_V = 0.4$\,mag,][]{Andrews:2011ApJ} would lead to lower inclinations of the stellar rotation axis, 
	which do not agree with the inclination of the disk.
	However, the disk inclination would be altered if the observed ellipses do not trace the rings' highes surfaces 
	above the midplane (called \textit{``edge ring 1/2''} in Figure~\ref{fig:flaring}).
	Without advanced radiative transfer modeling it is not possible to
	di\ff erentiate in our scattered light images between starlight scattered o\ffs the ring edges
	and light scattered o\ffs the slopes/walls of the rings (called \textit{``ring slope/wall''} in Figure~\ref{fig:flaring}).
	Consequently, the ring edges might truly
	lie further out than our ellipse fits.
	This e\ff ect would not be symmetric; we are more likely to have a larger contribution of the wall in the backward scattering side 
	(southwest, near the minor axis) of the rings.
	Therefore, if the apparent ellipses are a\ff ected by scattering by the ring walls, the ellipticity of the true ring edges (and their inclination) will be slightly lower and the ring o\ff sets larger.

\subsubsection{Nature of the ring structures}

	It is tempting to interpret the rings and arcs detected in the disk as spatial variations in surface density.
	However, our NIR scattered light detections of the disk trace the disk surface for the micron-sized dust grains. 
	We cannot \snew unambiguously \enew determine whether the rings are either a manifestation of variations in the scale height 
	caused by spatial variations in temperature (\snew e.g. due to \enew shocks), or variations in the surface density either caused by 
	dead zones or by massive planets carving a gap in the gas surface density.
	Dust trapping by local peaks in the gas pressure will appear di\ff erent for small and large dust grains.		
	Pinilla et al.~(submitted) show that if we can measure a di\ff erence between the density enhancements for the di\ff erent grain sizes, we can discriminate between dead zones creating a bump in the gas pressure and massive planets carving a gap in the gas disk.
	Furthermore, the mass of a gap carving planet can be predicted by the amount of the displacement between large grain and small grain peaks in the surface density for a given gas viscosity \citep{Juan:2016}.
	Large baseline sub-mm (ALMA) observations are therefore required to study the origin of the ring structures.
	A resolution comparable to our SPHERE observations will be needed in order to resolve the ring structure (when present for large grains), and accurately compare their radius with those of the rings in the surface of the small grain dust disk presented in this study.

	Another asymmetry is detected in the polarized intensity pro\fix le of Figure~\ref{fig:cut}, \snew 
	which roughly traces the major axes of the ellipses. \enew
	The peaks of $I1$ and $R2$ are brighter in the northwest than their southeastern counterparts, while
	the opposite (brighter in the southeast) is true for $A2$ and $R1$.
	The fact that these brightness asymmetries along the rings 
	are oscillating between northeast and southwest can possibly be explained with shadowing:
	the brighter parts of the rings might have a larger scale height than their fainter counterparts.
	If each ring outside of $I1$ is just marginally rising out from the shadow of the ring directly inside, 
	this would cause a brighter ring segment in the inner ring to cast a larger shadow on the outer ring, 
	\snew
	hindering the stellar radiation to heat up the 
	\enew
	segment in the outer ring with similar azimuth angle. 
	The opposite happens for the faint segments, which have a smaller scale height, casting less of a shadow on the next ring:
	this next ring receives more stellar radiation, heating it up and allowing it to `pu\ffs up' more, 
	making it brighter than the opposite side of the same ring.
	Shadowing of the outer rings by the inner rings is only possible when the \fl aring of the disk is 
	very small.
	When we compare the disk angle (parameter $d$ in Figure~\ref{fig:flaring}) for both rings, 
	we \fix nd that $\tan{d}(R1) = H_{\tau = 1}/r\,(R1) \approx 0.16$ 
	is marginally larger than $\tan{d}(R2) = H_{\tau = 1}/r\,(R2) \approx 0.15$, 
	which con\fix rms that the disk \fl aring is minimal.

\snew
	Although ALMA is detecting an increasing number of protoplanetary disks with multiple rings (e.g.~HL\,Tau),
	very few have been detected in scattered light.
	To the best of our knowledge, only TW\,Hya \citep[van Boekel et al.~submitted]{Rapson:2015}, HD\,141569A 
	\citep[e.g.~][]{Weinberger:1999, Perrot:2016} and HD\,97048 \citep{Ginski:2016} display multiple rings in scattered light.
	The inclination of HD\,141569A 
	(between $i = 45^\circ$ and $51^\circ$ for the di\ff erent rings, 
	\citealt{Biller:2015}, 
	and $i = 56^\circ$ for the entire disk, 
	\citealp{Mazoyer:2016}) is comparable to \rxjb.
	The ringed structure also looks very similar to \rxjb, because the rings are relatively sharp 
	compared to the larger radial extent of its gaps.
	As we discussed above, in a disk with low \fl aring (hereafter `\fl at', i.e.~$H/r =$ constant),
	a small ripple in the disk surface can cast large shadows outward.
	Indeed, \citet{Thi:2014} suggest that the disk of HD\,141569A is very \fl at,
	while the radial extent of rings in the surface of more \fl aring disks, such as HD\,97048 and TW\,Hya, is of similar size or larger
	than the width of their gaps.
	We therefore suggest that the sharpness of rings in the surface of a primordial disk 
	can be used as a tracer for the degree of \fl aring of the disk surface.
	This argument is based on the assumption that the apparent gaps are (mainly) due to 
	shadows cast by ripples in the scattering surface, 
	rather than true gaps in the surface density of the disk.
	To test this hypothesis we would need better knowledge of the scale height of the disk 
	(e.g. through high angular resolution gas observations with ALMA) 
	in several disks that show multiple ring structures in scattered light.

\enew
\subsubsection{$A1$: additional ring or bottom of $R1$}
\label{sec:ellipse2}
 \begin{table}[!h]
	\resizebox{1.\textwidth}{!}
	{\begin{minipage}{\textwidth}   	
	\begin{tabular}{ c|l|l|l|}
	  & Parameter            & Red & Purple \\
\hline \hline
	$A1$         & Semi-major axis ($''$)                      &   1.66  &   2.35   \\ 
	(no \fix t)   & Semi-minor axis ($''$)                 &   1.14  & 1.61 \\
		     & RA o\ff set $u_\mathrm{x}$ ($''$)     & 0.16  & -0.23 \\
		     & Dec o\ff set $u_\mathrm{y}$ ($''$)    & 0.11    & -0.16 \\
	             & O\ff set angle ($^\circ$)              &   56   & 236 \\
	             & $H_{\tau = 1}$ (au)                    &   49.2 & 70.7   \\
	             & $H_{\tau = 1}/r$                        &   0.16 & 0.16   \\
		\hline
	\end{tabular}
		\vspace{1mm}
\end{minipage}}
				\caption{
		Ellipse parameters for feature $A1$ in Figure~\ref{fig:irdisifs}c. 
		In Figure~\ref{fig:ellipses}, 
		the red ellipse shows the scenario where $A1$ is the bottom side of $R1$; the purple ellipse shows $A1$ as a separate outermost ellipse.
		}
		\label{tab:a1}
\end{table}
	\begin{figure*}	
		\centering
		\includegraphics[width=0.8\textwidth, trim = 80 0 80 0]{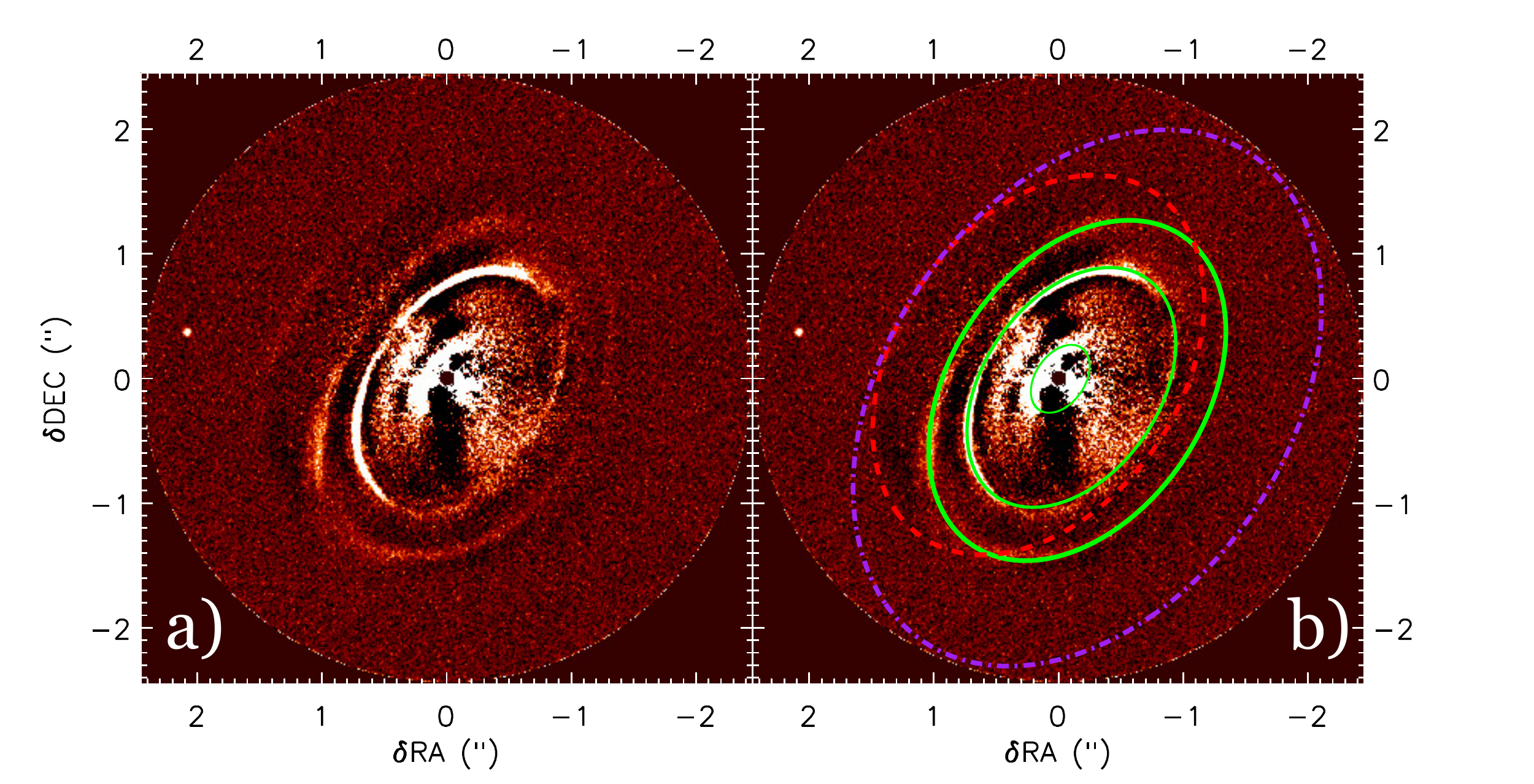}
		\includegraphics[width=0.79\textwidth, trim = 50 0 -10 20]{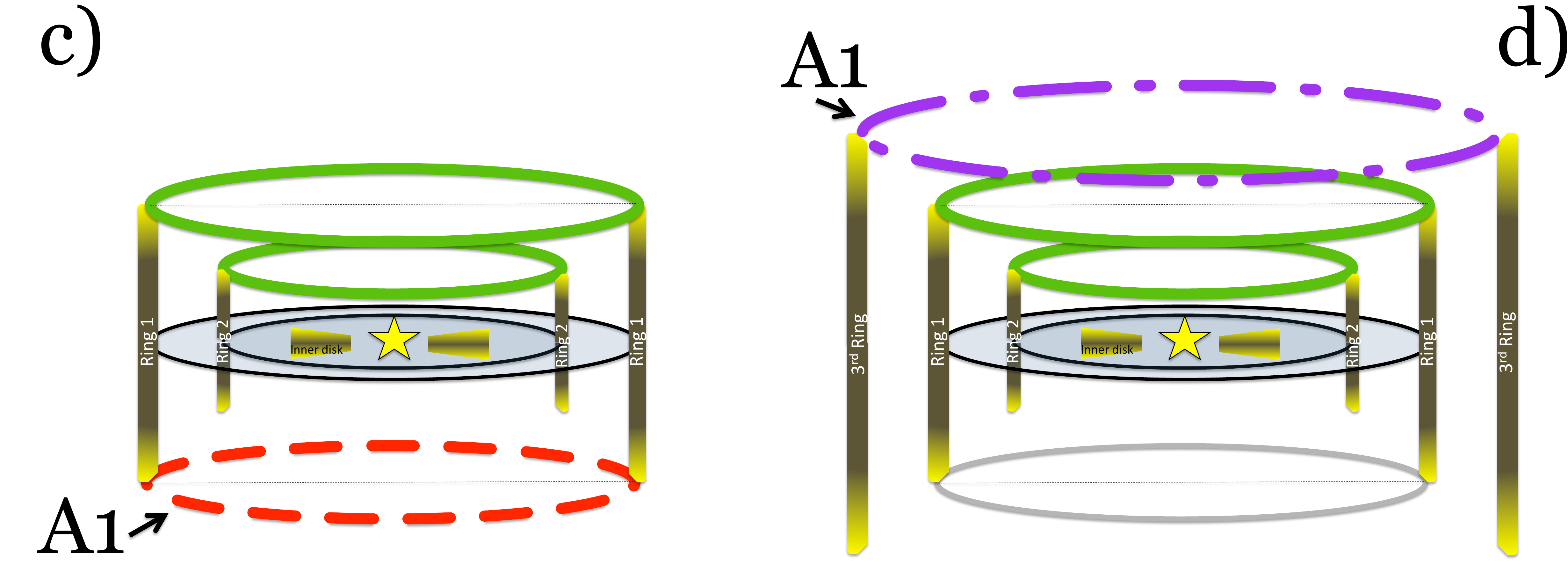}
		\caption{
		\textbf{b:} IRDIS $H23$-band TLOCI + ADI image.
			The overplotted green ellipses (solid lines) are the \fix ts to $R1$, $R2$ and $I1$, as listed in Table~\ref{tab:param}, 
			while the red (dashed) and purple (dash dot) ellipses depict the two scenarios where $A1$ is either the backward-facing side of ring 1, 
			or the forward-facing side of a 3rd ring further out
			(sketched in panels c and d), respectively. 
		\textbf{c:} First explanation for $A1$, for the same considerations as Figure~\ref{fig:flaring}: 
		$A1$ is the bottom/backward facing side of Ring 1, 
		for which $R1$ is the top/forward facing side, i.e. $r(A1) \approx r(R1)$.
		\textbf{d:} Second explanation for $A1$: $A1$ is the top/forward facing side of a third ring, with $r(A1) > r(R1)$.
		Both the red and the purple ellipses are selected based on the assumption that compared to $R1$ and $R2$,
		$A1$ would have roughly the same aspect ratio between major and minor axes as well as the same ratio between ellipse-star o\ff set and the major axis.
     		\label{fig:ellipses2}}
    	\end{figure*}

	For the arc-like structure ($A1$) in Figure~\ref{fig:irdisifs}, we consider two explanations to be equally plausible:
	it could either be an additional ring, at a separation from the star ($r$) larger than for $R1$. 
	However, the similarity of this arc to the shape of $R1$ at similar PAs ($10^\circ \lesssim PA \lesssim 100^\circ$) 
	is much stronger than when we compare $R1$ and $R2$ at these PAs.
	Therefore, we suggest an alternative explanation for $A1$ as it being the backward facing end of $R1$, 
	comparable to the bottom part of ring 1 in Figure~\ref{fig:flaring}. 

	To speculate on the shape of the ellipse for either explanation of ${A1}$, 
	we added two ellipses (red and purple) to Figure~\ref{fig:ellipses2}b. 
	The red and purple ellipses (explained in the cartoons of Figures~\ref{fig:ellipses}c and \ref{fig:ellipses}d) 
	are \fix xed at the intersection between ${A1}$ and the minor axis of the disk; have the same \textit{PA} as ${R2}$, 
	and the same ratio of major/minor axes (i.e. inclination).
	The \fix nal constraints 
	we used for the two ellipses is that 
	\begin{itemize}
		\item the ring has either a similar (or larger) $H_{\tau = 1}$ for the scenario where $A1$ is the backward facing side of $R1$ (red ellipse in Figure~\ref{fig:ellipses2}b and \ref{fig:ellipses2}c);
		\item $H_{\tau = 1}$ of a new ring further out (purple ellipse in Figure~\ref{fig:ellipses2}b and \ref{fig:ellipses2}d)
		needs to be large enough to rise from the shadow of $R1$.
		As we discussed in Section~\ref{sec:geo}, this means that $H_{\tau = 1}/r\,(A1) > H_{\tau = 1}/r\,(R1)$.
	\end{itemize}
	From Equation~\ref{eqn:ub} and the constraints above we derive that the absolute value of the ellipse (x-y) o\ff set ($|u|$) divided by the minor axis ($b$) for $A1$ should be:
	\begin{equation} 
		|u|/b\,(A1) \geq |u|/b\,(R1) = 0.18.
	\end{equation}
	We have used $|u|/b\,(A1) = 0.18$ for both ellipses in Figure~\ref{fig:ellipses2}b, which can be considered 
	as a best guess for the red ellipse and a lower limit for the purple ellipse: larger purple ellipses (which will have larger $|u|/b$) are not ruled out.
	Deeper observations of this structure will reveal which of the two scenarios is true: 
	detecting a larger (azimuth coverage of the) ring segment will enable us to distinguish between the
red and the purple ellipse scenarios.
	
	In order to \fix x the red ellipse on the intersection between the minor axis and the detected $A1$ feature,  
	both a larger semi-major axis and X and Y o\ff set (in opposite direction) are needed for the red ellipse (see Table~\ref{tab:a1}) 
	than what we found for the green ellipse of $R1$ in Table~\ref{tab:param}.
	This either hints at a structure with $H_{\tau = 1,bottom} > H_{\tau = 1,top}$ or $r\,(A1) > r\,(R1)$. 
	The scattering-angle for the backward facing part of ring 1 will be di\ff erent than the forward facing part: 
	moving the beam back into denser regions of the disk instead of away from higher density (as for the forward facing $R1$).
	This will bring the backward facing $\tau = 1$ surface to lie further from the midplane than for the forward facing surface at the same distance $r$.

\subsection{Possible disk sculpting companion}

Should the ring ${A1}$ turn out to be the backward facing side of ring 1, as we illustrate in Figure~\ref{fig:ellipses2}c, 
we would expect there to be a fairly massive planetary companion to the system. 
To make the backward facing side of ring 1 visible, the disk needs to be truncated at most 
$H(A1) \times \tan{i} = 54$\,au beyond ring 1, which lies at $1.66 \times 185 = 307$\,au 
(where we used the semi-major axis of $1.66''$ for $A1$, 
instead of the $1.50''$ for $R1$, since the former will be dominated by the outer edge, the latter by the inner edge of ring 1).
The disk truncation at $r \approx 361$\,au can be a consequence of a massive planet beyond this radius, opening  a gap in the disk. 
Detection of such a companion should be possible 
with our IRDIS ADI observations as long as the planet signal is not obscured by the disk (e.g. when the planet is on the western side of its orbit, or the truncation is not performed by multiple lower mass planets.
Should companion candidate 1 (cc1, see Figure~\ref{fig:irdisifs}) be associated with the system it could provide the required disk truncation. 

To determine the mass of cc1 (if it is indeed bound to the system) we \fix rst need to \fix nd the age of the system.
In Appendix~\ref{sec:stellar2} we determine the rotation period ($P = 5.72\pm0.01$\,d), spectral type (K5-K6) 
and temperature ($T_\mathrm{e\ff}=4100\pm100$\,K) of \rxjb,
which we use to determine the inclination of the stellar rotation axis and constrain the stellar age and mass.
Overall, the properties of the star such as accretion, characteristics of the
disk, the kinematic, and the limits from lithium and rotation period indicate an age less than 5\,Myr.
Following the same process as \citet{Wahhaj:2010ApJ} we use the models of \citet{Siess:2000A&A} to determine the age of \rxj from isochrones. Adopting the luminosity $L=1.07 \pm 0.11 L_\odot$ and e\ff ective temperature $T_\mathrm{e\ff}=4100 \pm 100$, as stated before, we \fix nd age range of $1.8\pm0.6$\,Myr and a mass of $0.8 \pm 0.1$\,M$_{\odot}$. 

Using the COND model \citep{Baraffe:2003}, our observed $H$-band magnitude, and the assumed age of the system,
we estimate the mass of cc1 to be $0.8\pm 0.1$\,M$_\mathrm{jup}$. 
In this estimation, we have assumed the signal to be dominated by thermal emission of the planet, rather than the emission by a circumplanetary accretion disk. If we assume that cc1 is a companion orbiting \rxj along the midplane of the disk, its deprojected distance to the star would be $\sim 540$\,au. 
Assuming that the maximum gap size created by a companion on a circular orbit is $\sim 5$ Hill Radii \citep{DodsonR:2011, Pinilla:2012},
at this separation we would need a planet with higher mass ($\gtrsim 1.5$\,M$_\mathrm{jup}$) to fully truncate the gas disk at $r =361$\,au,
or the planet needs to have an elliptical orbit.
However, if the gas density is su\ff \mbox{}iciently reduced in the outer disk, in particular at the planet gap, the $\mu$m sized grains can become decoupled to start drifting inwards, leading to a smaller disk truncation of the micron-sized particles compared to the gas.  
Without the presence of a planetary gap, the micron sized particles would remain coupled and the truncation in scattered NIR light would be further out than observed. 
Whether the potential drift of the $\mu$m grains in a continuous gas disk su\ff \mbox{}ices to reduce the disk optical depth enough to enable the transmission of light scattered by the backward-facing side of ring 1 requires radiative transfer modelling, which is beyond the scope of this study. \\

\begin{figure}
   		\centering
		\includegraphics[width=0.5\textwidth, trim = 20 0  0 -5]{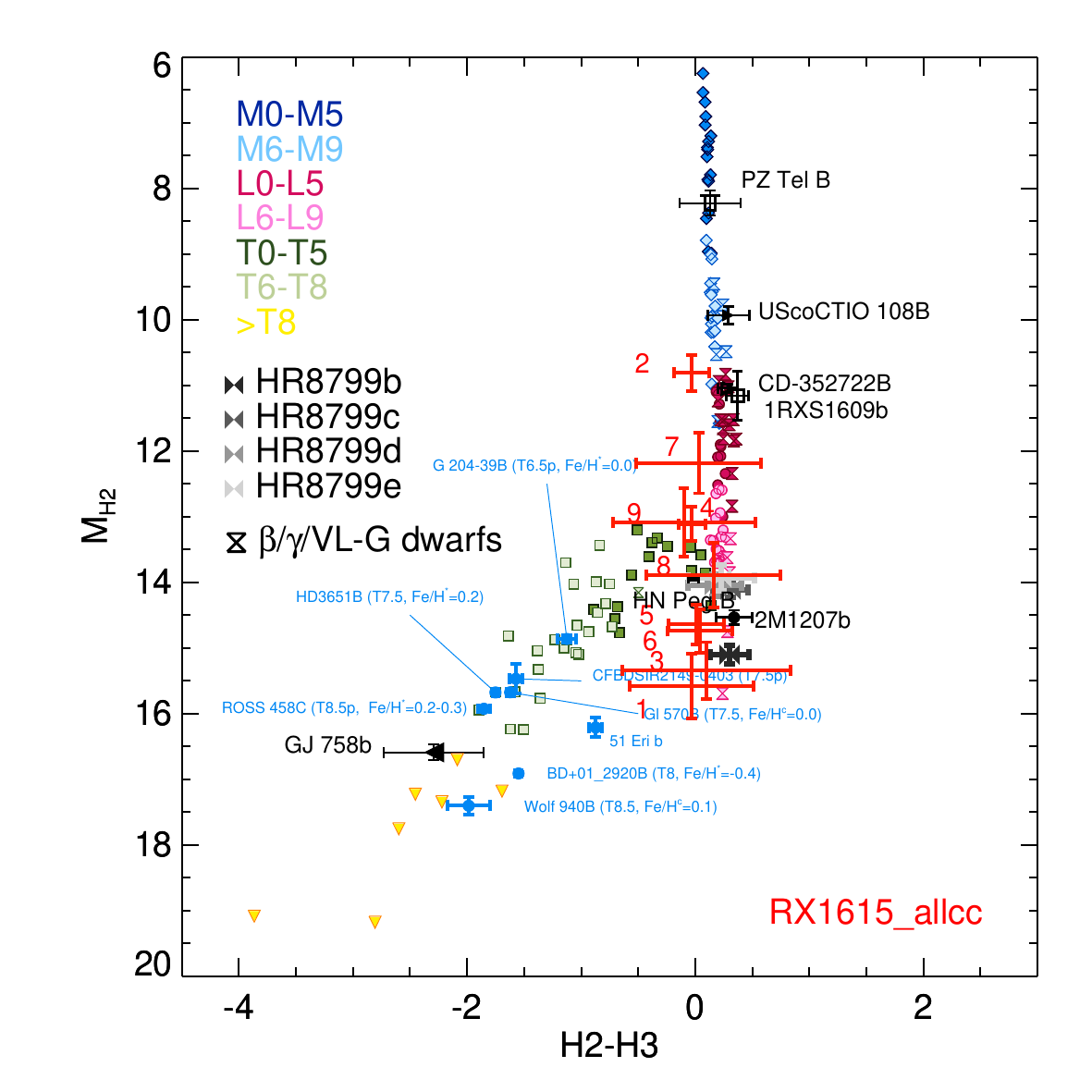}
   		   	\caption{Color magnitude diagram displaying our candidate companions compared to already known planetary companions and brown dwarfs. Candidate companions are displayed, and numbered, in red with corresponding error bars.
   		   	}
      		\label{fig:cmd}
      	\end{figure}

We show in Figure~\ref{fig:cmd} the locations of our candidate companions on a color magnitude diagram, and included previously known brown dwarfs and planetary companions for reference. 
Upon first inspection of this diagram we might assume that cc1 is more likely to be a background object as it, along with all remaining candidates, has an $H2 - H3$ color typical of a background M dwarf. 
However, we know that some planetary companions (such as those of HR8799 and 2M1707) do have colors within the range of cc1 so we can claim nothing for certain. 
Indeed \rxj is treading on new territory being so young and low mass. If cc1 is attributed to the \rxj system 
we would normally expect there to be a methane signature (which would yield a negative $H2-H3$ color). 
However, to date no observational evidence for methane emission in such young and low mass systems exists,
leaving the possibility open that cc1 is bound to the \rxj system.
Follow-up observations of cc1 are needed to con\fix rm whether the candidate is co-moving with \rxjb.
If bound cc1 would be an ideal target for characterization with the IRDIS low resolution long-slit spectroscopy mode \citep{Vigan:2008}.

\begin{figure}
   		\centering
		\includegraphics[width=0.5\textwidth, trim = 20 10  5 -5]{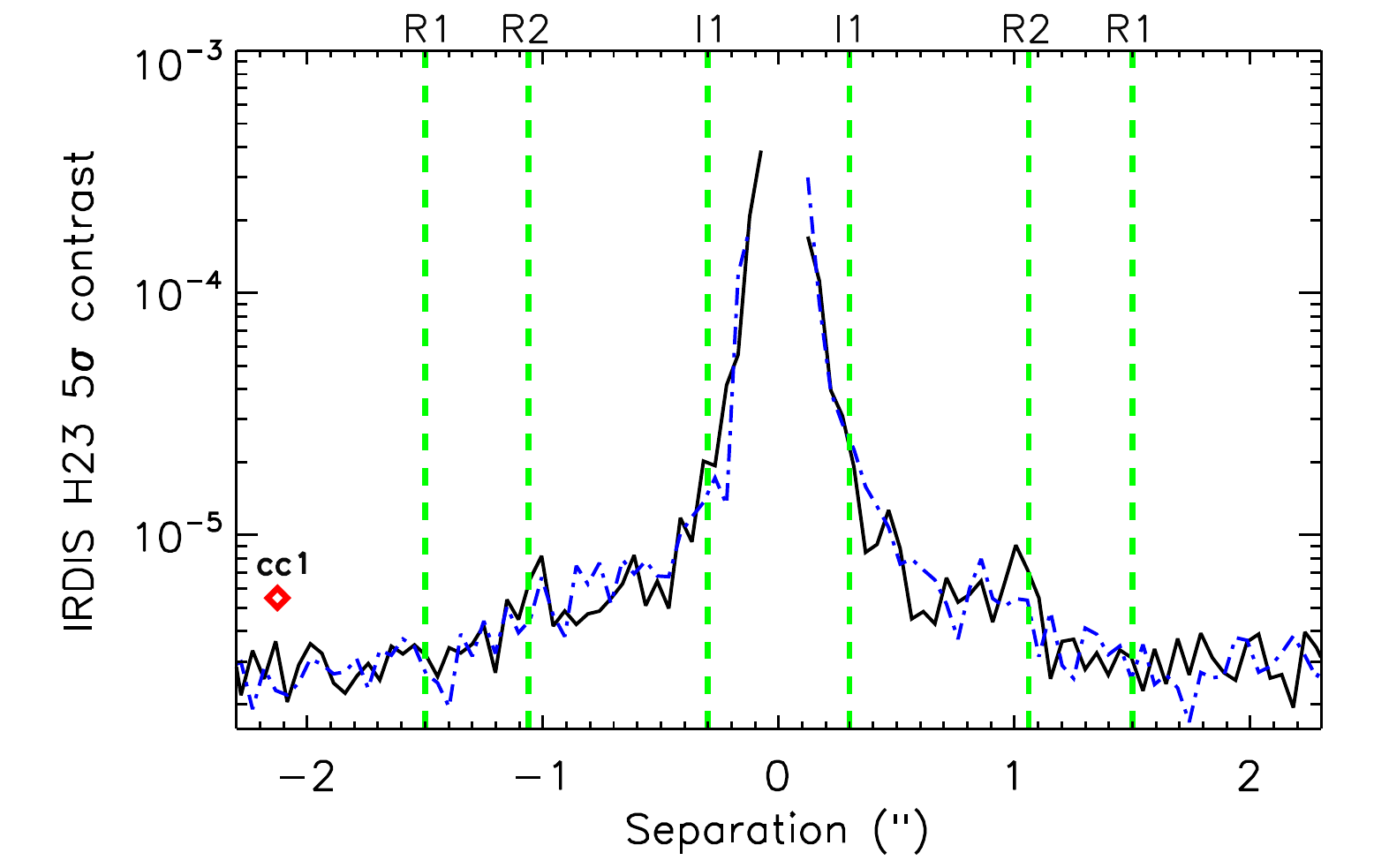}
   		   	\caption{The $5 \sigma$ contrast measured in the H23 TLOCI reduction of May 15 2015, 
   		   	along \textit{PA} $ = 146^\circ$, parallel to the ellipse's major axes 
   		   	(black solid line, negative is southeast);
   		   	and \textit{PA} $ = 56^\circ$, the ellipse's minor axes (blue dash-dotted line, negative is northeast).
   		   	The vertical green dashed lines display the semi-major axes of $R1$, $R2$, and $I1$ 
   		   	as annotated above the top axis of the plot. 
   		   	The red diamond shows the measured signal of cc1 (which, we \fix nd, corresponds to 
   		   	$0.8 \pm 0.1M_\mathrm{jup}$). 
   		   	Notice that the actual location of cc1 does not coincide with either the minor or the major axes of the disk.
   		   	}
      		\label{fig:contrast}
      	\end{figure}

Besides the option of the disk being truncated by a single massive planet as an explanation of why $A1$ could possibly be the backward facing side of $R1$,
multiple lower mass planets can provide a similar truncation of the disk. 
Such lower mass planets might remain undetectabel by SPHERE.
For this reason we chose to include the contrast plot seen in Figure~\ref{fig:contrast}, showing the contrast in two cuts across the center of \rxjb, parallel to the major and along minor axes of the rings.
A conventional contrast curve, where the contrast is measured over concentric annuli would contain the disk signal in such a way that one can no longer tell if a peak is caused by a ring or by something else.
However, outwards of R1 this is not an issue. We have inserted the measured signal of cc1 at its appropriate radial separation for reference.

\subsection{Wavelength dependent surface brightness of $R2$}

	$R2$ is clearly detected in $H23$-band ADI image; it is fainter in the $J$-band PDI data; 
	and can no longer be distinguished against the background disk signal in the $R'$-band PDI image.
	We detect a similar wavelength dependence of the $R2$ surface brightness when we compare the 
	three wavelength regimes of the IFS in Figure~\ref{fig:ifs}.
	If this wavelength dependence is truely astrophysical, 
	it could be indicative of spatial/radial variations in the chemistry or grain size distribution throughout the disk.
	However, the low Strehl ratio and FWHM of the ZIMPOL data described in Section~\ref{sec:obsred} 
	can also be responsible for washing out a structure which is not resolved in the radial direction, 
	smoothing out the unresolved structure so much that we can no longer distinguish $R2$ from the surface brightness of its surroundings.
	The simultaneous measurements at the three IFS wavelength regimes suffer from the same systematic e\ff ect:
	the Strehl ratio increases for longer wavelength. 

	Since disk signal is present both directly inside and outside of $R2$, we cannot determine if the low resolution of the $R'$ image has washed out the disk signal. 
	Convolving a radiative transfer model with the PSFs of the di\ff erent observations will be the best way to answer whether we are looking at a 
	true astrophysical change with wavelength of the surface brightness or rather at systematic e\ff ects due to the di\ff erence in Strehl ratio and FWHM. 
	However, creating a realistic radiative transfer model lies outside the scope of this paper and is left for future work.

\section{Conclusion}
\label{sec:conclude}

We have studied the system of \rxj with four observing modes of the high contrast imager VLT/SPHERE.
We detect the disk of \rxj in scattered light for the \fix rst time, 
from the optical $R'$-band to the NIR $K_s$-band, with
all high contrast imaging modes used in this study.
The $J$ and $H23$ band images show three elliptical disk components surrounding the star and two arc-like features.
The two outer ellipses $R1$ and $R2$ have their centers o\ff set from the position of the star.
The simplest explanation for the elliptical features is that we see a disk at inclination $i = 47 \pm 2^\circ$ which
contains an inner disk $I1$ 
surrounded by two circular rings, for which the height of the $\tau = 1$ surface increases with its distance to the star.
Ellipse \fix tting yields the major axes (i.e. radii (r) of the rings) for $R1$ to be $278 \pm 2$\,au, 
for $R2$ $r = 196 \pm 2$\,au, and for $I1$ $r = 56 \pm 2$\,au.
A tentative cavity is detected inside the inner disk.
The apparent o\ff sets between star and ring centers allow us to determine 
the height of the scattering surface above the midplane: $H_{\tau = 1}(R1) = 44.8 \pm 2.3$\,au and 
$H_{\tau = 1}(R2) = 30.2 \pm 2.4$\,au.

For the outermost disk feature ${A1}$, our detection was not deep enough to determine its origin.
Deeper observations are needed to determine whether $A1$ is the backward facing side of $R1$ 
or a new ring at larger separation from the star.
ALMA observations might allow us to directly image the thickness of the gas disk, as was done for 
HD163296 by de \citet{Gregorio:2013A&A}. 
If we detect the dust disk truncated between $r = 300 - 360$\,au, and the gas disk with $H$ comparable to the thickness we derived, this would be a strong con\fix rmation of our understanding of the disk geometry.

Nine companion candidates are detected between 2.1 - 8.0$''$ with SPHERE/IRDIS.
In VLT/NACO data we detect four of these nine point sources, and determine that they are not co-moving, 
and therefore not bound to the system. 
Follow-up observations of the remaining \fix ve viable companion candidates need to be made to determine if they are co-moving.
If cc1 indeed turns out to be orbiting the disk at $540$\,au and if 
$A1$ is indeed showing the backward facing side of a truncated disk,
\rxj provides the most unambiguous example of ongoing planet-disk interaction yet detected.

\begin{acknowledgements}
We thank the anonymous referee for his/her very rapid and constructive comments.
Many thanks go out to the instrument scientists and operators of the ESO Paranal observatory for their 
support during the observations.
We also thank Michiel Min for the insightful discussion about the geometry of the disk.
P. Pinilla is supported by a Royal Netherlands Academy of Arts and Sciences (KNAW) professor prize.
AJ is supported by the DISCSIM project, grant agreement 341137 funded by the European Research Council under ERC-2013-ADG.
MM and CG acknowledge the German science foundation for support in the programme MU2695/13-1.
This research has made use of the SIMBAD database, operated at the CDS, Strasbourg, France.
NASA’s Astrophysics Data System Bibliographic Services has been very useful for this research.
SPHERE is an instrument designed and built by a consortium consisting of IPAG (Grenoble, France), MPIA (Heidelberg, Germany), 
LAM (Marseille, France), LESIA (Paris, France), Laboratoire Lagrange (Nice, France), INAF - Osservatorio di Padova (Italy), 
Observatoire de Gen\`eve (Switzerland), ETH Zurich (Switzerland), NOVA (Netherlands), ONERA (France), and ASTRON (Netherlands) 
in collaboration with ESO. 
SPHERE was funded by ESO, with additional contributions from the CNRS (France), MPIA (Germany), INAF (Italy), FINES (Switzerland) and 
NOVA (Netherlands). 
SPHERE also received funding from the European Commission Sixth and Seventh Framework Programs as part of the 
Optical Infrared Coordination Network for Astronomy (OPTICON) under grant 
number RII3-Ct-2004-001566 for FP6 (2004-2008), 
grant number 226604 for FP7 (2009-2012), and grant number 312430 for FP7 (2013-2016).
\end{acknowledgements}

\bibliography{ref160413}   
\bibliographystyle{aa}  

\appendix
\section{Stellar Properties}
\label{sec:stellar2}

\subsubsection*{Rotation Period}
We have taken advantage of publicly available SuperWASP 
\citep{Butters:2010A&A} photometry time series data, collected during 2004 - 2006, 
to determine the rotation period. 
This consists of 12684 $V$-band measurements with an average photometric precision $\sigma = 0.018$\,mag. 
After the removal of outliers and low-precision measurements from the time series by applying a
moving boxcar \fix lter with 3$\sigma$ threshold, we average consecutive data collected within 1\,hr.
Finally we are left with 545 averaged magnitudes with an associated standard
deviation $\sigma = 0.010$\,mag for the subsequent analysis. 
\snew
To search for the rotation period of \rxjb, we apply the Lomb-Scargle method \citep[LS,][]{Scargle82},
with the prescription of  \citet{Horne:1986ApJ}, as well as the Clean \citep{Roberts87} periodogram analyses of the data.
\enew
In all three SuperWASP observing seasons (2004, 2005 and 2006) we have searched in the period range 0.1 - 100\,d, 
and \fix nd with both LS and Clean the same rotation period $P = 5.719\pm 0.014$\,d, which is
the \fix rst determination for the rotation period of \rxjb. 
Although other peaks due to observation timings and beat frequencies are present in the periodogram,
we do not detect any other signi\fix cant peaks of interest.

\rxj was also observed by the All Sky Automated Survey \citep[ASAS,][]{Pojmanski:1997AcA} in the years 2001-2009. 
Despite the lower photometric precision, our LS and Clean analyses allowed us to
measure the same $P = 5.72\pm0.01$\,d rotation period and 
a peak-to-peak light curve amplitude
of $\Delta V = 0.12$\,mag.

	\begin{figure*}	
		\centering
		\includegraphics[width=0.9\textwidth, trim = 70 0 0 20]{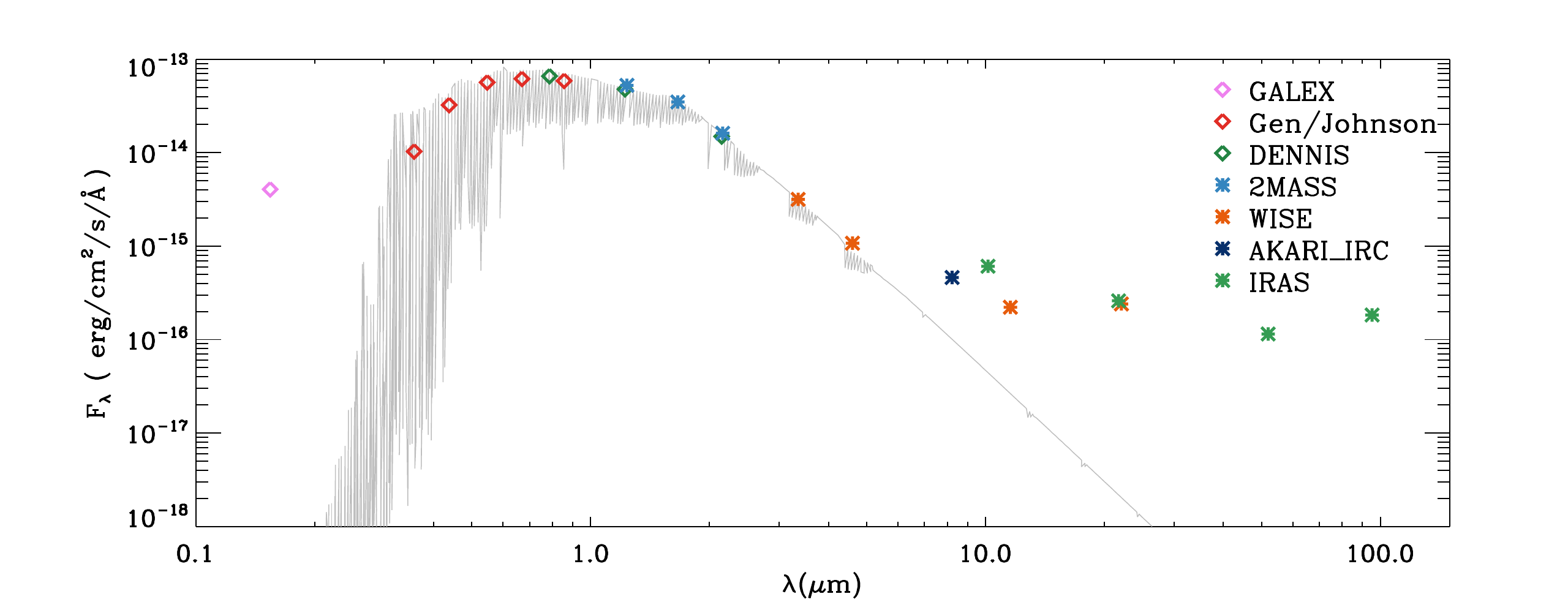}
		\caption{
	Spectral Energy Distributions (SED, see legend) and  our best \fix t BT-NextGen model (grey solid line, $T_\mathrm{e\ff} = 4100$\,K) for \rxjb.
     		\label{fig:SED}}
    	\end{figure*}

\subsubsection*{Spectral Type and Temperature}
In the literature the spectral type is found in the range from K4  \citep[based on Spitzer/IRS spectra]{Merin:2010ApJ}
to K7  \citep[using spectroscopy with VLT/X-Shooter]{Manara:2014A&A}, with \citet[with visible and NIR photometry]{Wichmann99} and \citet[based on visible light spectroscopy]{Krautter:1997A&AS}
favouring a K5 classi\fix cation.
Using the sequence of intrinsic colors and temperatures of pre-main sequence stars by  \citet{Pecaut13} 
and comparing it to the measured colors by  \citet{Makarov:2007ApJ} 
we \fix nd that the photometric colors are fully compatible with a K5-K6 star suffering a small or even negligible amount of reddening. 
The colors of a K4 star would instead indicate a reddening of about E(B$-$V) = 0.15\,mag 
while a K7 star is expected to have redder colors than those observed. 
Previous estimates of \snew extinction \enew span between
$A_V = 0$\,mag \citep{Manara:2014A&A} to $A_V = 1.1$\,mag \citep{Wahhaj:2010ApJ} with an intermediate value
$A_V = 0.4$\,mag by \citet{Andrews:2011ApJ}.

Assuming the spectral type K5-K6 and a low amount of reddening (E(B$-$V) $< 0.15$\,mag), the corresponding
e\ff ective temperatures in the scale by  \citet{Pecaut13} are $T_\mathrm{e\ff} = 4140-4020$\,K for K5 and
K6 spectral types respectively.
We have also used the available optical, near-IR, and IR photometry to build the observed SED (Figure~\ref{fig:SED}). 
\snew
We have obtained the Far Ultra Violet magnitudes from the GALEX catalogs \citep{Bianchi:2011};
\enew
the UBVRI magnitudes are taken from  \citet{Makarov:2007ApJ}; 
Denis $IJK$ magnitudes from DENIS Catalogue  \citep{Ochsenbein00}; $JHK$
magnitudes from the 2MASS project \citep{Cutri:2003}; W1-W4 magnitudes from the
WISE project \citep{Cutri:2013}; IRAS magnitudes from the IRAS Catalog of Point
Sources \citep{Helou88}; AKARI magnitudes from AKARI/IRC mid-IR all-sky
Survey \citep{Ishihara10}. 
The SED was \fix t with a grid of theoretical spectra from
the BT-NextGen Model \citep{Allard:2012} 
and the best \fix t, shown in Figure~\ref{fig:SED} is obtained with a model of
$T_\mathrm{e\ff} = 4100\pm50$\,K, when adopting a distance $d = 185$\,pc and E(B$-$V) = 0.00\,mag. 
We note the presence of a signi\fix cant UV and IR excess which is common when there is an associated disk. 
Combining our two estimates we conservatively take the temperature to be $T_\mathrm{e\ff}=4100\pm100$\,K.

\subsubsection*{Stellar rotation axis inclination}
Using the brightest visual magnitude $V = 12.0$\,mag
from the ASAS time series, the distance $d = 185$\,pc, $A_V = 0.00$\,mag, bolometric correction
from \citet{Pecaut13} tables, we derive the stellar luminosity $L_\star = 1.07\pm0.11\,L_{\odot}$. 
For $T_\mathrm{e\ff} = 4100$\,K, we derive the stellar radius $R_\star = 2.05 \pm 0.21\,R_{\odot}$. 
With the stellar radius and the rotation period determined above 
we can compute the rotational velocity at the stellar surface $v = 18.1$\,km\,s$^{-1}$.
From the projected rotational velocity
$v \sin{i} = 13.0$\,km\,s$^{-1}$ 
\citep{Wichmann99}, we derive the inclination of the rotation axis $i = 46 \pm 6^\circ$, 
which is compatible within errors with the inclination $i = 45 \pm 5^\circ$ of the disk
derived by \citet[\& private communication]{Marel:2015A&A}. 
Therefore, we infer that the disk is coplanar with the stellar equatorial plane.
If we instead assume an \snew extinction \enew $A_V = 0.4$\,mag \citep{Andrews:2011ApJ} to determine $L_\star$ and $R_\star$, 
the inferred inclination $i = 37 \pm 6^{\circ}$. 

\subsubsection*{Age, mass and distance}
A comparison of the rotation period of \rxj with the distribution of rotation periods of stellar associations of known age helps us to constrain the age. 
We \fix nd that the rotation period $P = 5.719$\,d is longer than for members of the TW Hya association \citep[$P = 4$\,d, age = 9\,Myr,][]{Messina:2010}.
Considering that \rxj is a pre-main sequence star whose radius is still contracting and its rotation still spinning up, we can state that \rxj is not older than about 9\,Myr.
The measured stellar lithium (Li) equivalent width of $558$\,m\AA~\citep{Wichmann99}
\fix ts very well into the Li distribution of the 5\,Myr NGC 2264 open cluster (Bouvier et al., submitted), 
leading us to expect an even younger age compatible with that expected for a CTTS with a gas-rich disk.

\citet{Makarov:2007ApJ} classi\fix ed the star as a member of the Lupus association, deriving a kinematic
distance of 184\,pc. They also noted a group of stars that appear younger ($\sim 1$\,Myr as opposed to $\sim 5$\,Myr) and at a greater distance than the bulk of the members of the association. However, the star is not considered in the most recent study of Lupus \citep{Galli13}.

\end{document}